\documentclass{aa}
\usepackage{graphicx}
\usepackage{txfonts}
\usepackage{natbib}
\usepackage{longtable}
\usepackage[shortlabels]{enumitem}
\usepackage{siunitx}
\usepackage{epstopdf}
\usepackage{mathrsfs}

\usepackage{hyperref}	
\hypersetup{colorlinks=true,linkcolor=blue,citecolor=blue,filecolor=blue,urlcolor=blue}
\usepackage{enumerate}%
\newcommand{\civ}{\ifmmode {\rm C}\,\textsc{iv} \else C\,\textsc{iv}\fi}

\begin{document}

\title{On the connection between galactic downsizing and the most fundamental galactic scaling relations}
\author { E. Spitoni\inst{1}  \thanks {email to: spitoni@phys.au.dk}
  \and F. Calura
  \inst{2} \and M. Mignoli\inst{2}  \and R. Gilli\inst{2} \and
  V. Silva  Aguirre\inst{1} \and A. Gallazzi \inst{3} }
\institute{Stellar Astrophysics Centre, Department of Physics and
  Astronomy, Aarhus University, Ny Munkegade 120, DK-8000 Aarhus C,
  Denmark \and INAF-OAS, Osservatorio di Astrofisica e Scienza dello Spazio di Bologna, via Gobetti 93/3, I-40129 Bologna, Italy \and INAF – Osservatorio Astrofisico di Arcetri, Largo Enrico Fermi 5, I-50125 Firenze, Italy}

\date{Received xxxx / Accepted xxxx}

\abstract
    {In their evolution, star-forming galaxies are known to follow scaling relations between some fundamental physical quantities,
      such as the mass-metallicity  and the star formation main sequence relations.   }
{We aim at studying the evolution of galaxies that, at a given redshift, lie simultaneously on the mass-metallicity  and main sequence relations (MZR, MSR). }
{To this aim, we use the analytical, 'leaky-box' chemical evolution model  of \citet{spitoni2017}, in which galaxy evolution is
described by infall timescale $\tau$ and  wind efficiency $\lambda$. 
We provide a  detailed analysis of  the temporal evolution of their metallicity, stellar mass, mass weighted age  and gas fraction.}
{The evolution of the galaxies lying on the MZR and MSR at $z\sim0.1$ suggests that the average infall time-scale 
in  two  different  bins  of  stellar  masses ($M_{\star}<10^{10} M_{\odot}$
and $M_{\star}>10^{10} M_{\odot}$)  decreases with
decreasing redshift  through the addition of new galaxies with shorter timescales. This means that at each redshift, only the
youngest galaxies can be assembled on the shortest timescales and 
still belong to the star-forming MSR. In the lowest mass bin, a
decrease of the median $\tau$ is accompanied by an increase of the
median $\lambda$ value.  This implies that systems which have formed
at more recent times will need to eject a larger amount of mass to
keep their metallicity at low values. Another important result is that galactic downsizing, as traced by the age-mass relation, is naturally recovered by imposing the local MZR and MSR for star-forming galaxies.
This result is retained even assuming a constant star formation efficiency for different galactic masses (without imposing the observed scaling relation  between stellar mass and gas depletion time-scales).
Finally, we study the evolution of the hosts of \civ-selected AGN, which at $z\sim 2$ follow a flat MZR, as found by \citet{mignoli2019}.
If we impose that these systems lie on the MSR, at lower redshifts we find an 'inverted' MZR, meaning that some additional processes
must be at play in their evolution.}
{In our model, galactic downsizing is a direct consequence of the mass-metallicity and main sequence relations for star-forming galaxies.
 This poses a challenge for models of galaxy evolution within a cosmological framework.}


\titlerunning{Downsizing and scaling relations}

\authorrunning{Spitoni et al.}

\maketitle

\section{Introduction}\label{introduction}

In the last decades, several observational campaigns have confirmed
that the evolution of star-forming galaxies is described by scaling
relations  between some fundamental physical quantities. For instance,
a tight relation has been found between galactic stellar mass and
metallicity  \citep{lequeux1979,maiolino2019}, the so-called mass-metallicity
relation (hereinafter MZR).  In the local Universe,
the assessment of the MZR and other scaling relations 
was possible thanks to high-quality data 
collected within the Sloan Digital
Sky  Survey (SDSS), which have enabled to 
measure the
flux ratios of the main optical emission lines for more than 100000
galaxies (e.g., \citealt{tremonti2004,mannucci2010,perez2013,lian2015}). 

Moreover, the observed evolution with redshift of the MZR 
(i.e. \citealt{maiolino2008,yuan2013,zahid2014}) 
allows one to trace 
the chemical enrichment history of galaxies throughout different cosmic epochs. 
\citet{zahid2014} pointed out that at redshifts $z<1.6$, the MZR
follows a steep slope with a `knee'  at  a characteristic turnover
mass at $ M_{*} \sim 10^{10}~M_{\odot}$.
At stellar mass values higher than that, the MZR flattens as in the most massive galaxies 
metallicities begin to saturate.
As a consequence, the redshift evolution of the
MZR depends only on the evolution of the
characteristic turnover mass. 
 The evolving turnover mass can be related to the change in the gas-to-stellar mass ratio
  (see e.g. also the similar dependence between turnover mass and SFR in \citealt{curti2020}).

Different theoretical scenarios have been presented in the past in order to explain the
MZR, and they can be summarized as follows:

\begin{enumerate}[i)]

 \item 
Galactic outflows: low-mass galaxies are more
efficient in expelling metal-enriched matter
than giant galaxies, because of the shallower gravitational
potential wells of the former 
\citep{larson1974,tremonti2004,spitoni2010,spitoni2017,hirschmann2016}.
In the framework of pure chemical evolution models,
a more recent work of \citet{lian2018} suggested that 
strong metal outflows occurring in the earliest galactic evolutionary phases
also help reconciliating gas phase and the mass-weighted stellar MZRs.

\item Variable star formation efficiency (SFE): 
the efficiency of star formation is larger in more massive
systems, which have formed the bulk of their 
stars by means of an intense star formation event at high redshift,
quickly enriching their ISM to solar or over-solar metallicities. 

In the local universe,  \citet{boselli2014} presented a scaling
relation between the typical galaxy gas depletion timescale  and the
galaxy stellar mass that can be easily transformed in SFE.
They observed that massive galaxies consume the available gas
reservoir on shorter typical timescales than dwarf galaxies. 
This means that
larger galaxies are expected to experience, on average, higher SFEs
(see also \citealt{matteucci2012}).
Various chemical evolution 
studies support this scenario \citep{matteucci1994, 
  lequeux1979,caluraMZ2009}.
\item A third interpretation of the MZ relation is linked to the initial stellar mass function. \citep{koppen2007}
showed how the MZ relation can be explained by a higher upper mass cutoff in the initial mass function (IMF) in more massive galaxies.  
\end{enumerate}

Chemical evolution models are powerful tools which help to probe fundamental processes regulating 
galaxy formation and evolution. 
They can provide important constraints on how subsequent stellar
 generations have  modified  the chemical composition of the
 interstellar medium (ISM), to give place to 
 chemical abundance pattern as observed in present-day galaxies \citep{matteucci2012}.  
In this framework, \citet{spitoni2017} presented  an analytical solution for the
evolution of the galaxy metallicity, where 
an  exponential infall of 
gas was assumed for the gas accretion rate, 
along with instantaneous mixing and instantaneous recycling
approximation (IRA) \citep{matteucci2012}.

In their study of the MZR in the local SDSS  star-forming galaxies by \citet{peng2015},
\citet{spitoni2017} found that lower mass galaxies suffered more intense winds compared to the higher mass ones,
in agreement with the scenario i) mentioned above,  and had to be characterized by shorter time-scales of gas accretion
imposing the variable SFE by \citet{boselli2014}, as indicated by the scenario ii). 

Another fundamental quantity to understand galaxy evolution is
the star formation rate (SFR), which helps 
to shed light on the basic processes regulating the 
conversion of gas into stars and the growth of the stellar mass.
The observations of large samples of star-forming galaxies in different redshift intervals 
 have allowed to establish the existence of
a well-defined relation between the SFR and the stellar mass, 
the so-called Main Sequence relation (MSR
hereafter), thoroughly studied in several work 
 \citep{brinchmann2004,elbaz2007,noeske2007,peng2010,santini2017,pearson2018}.
As the MZR, also the MSR is characterized by a clear evolution with redshift. 
 \citet{pearson2018} studied the  MSR in the redshift range
between  $z=0.2-6$, showing that the slope of this
relation does not change substantially with cosmic time. 
In particular, the slope of the MSR becomes
steeper only for galaxies at high redshift,  in the range between
$3.8\le z \le 6$.  
On the other hand, it has been ascertained in several studies that 
the zero point of the MSR increases with redshift (e. g., \citealt{santini2017}, \citealt{iyer18}). 

One of the most remarkable outcome of studies of scaling relations in local
and distant galaxies was the discovery of substantial differences in the star formation history of
low- and high-mass galaxies, also known as ‘galaxy downsizing’ (DS) \citep[e.g.][]{cowie1996, mortlock2011}.
In the framework of galactic DS, the largest galaxies are
assembled at the earliest  times, with a large SFR values mostly
concentrated in their earliest phases and accompanied by very 
efficient chemical enrichment 
\citep{caluraMZ2009,matteucci2012,maiolino2019}. 
The phenomenology of Galactic downsizing is wide and multifaceted, as it is reflected by
several observational aspects of galaxies at both low and high redshift,
including color distributions at optical and infrared wavelegths of high-redshift galaxies \citep{cowie1996},
studies of the integrated abundance rations in local early type galaxies (e. g., \citealt{spolaor10} and references therein),
studies of the evolution of the galactic stellar mass function (e. g., \citealt{mortlock11}) and studies of metallicity, age
and stellar mass from optical spectra of local galaxies (\citealt{gallazzi05}) 

Altogether, these independent pieces provide the same indications, i. e. 
a more intense star formation activity in the most massive galaxies at early epochs, 
followed by higher star formation  in low-mass galaxies at lower redshifts.

By means of an analytic chemical evolution model of \citet{spitoni2017},  
our aim is to study the the evolution of some basic properties of the galaxies which build the MZR and
MSR at different redshifts. These properties include stellar mass and star formation activity, and how
these are related to other quantities, such as the ages of the stellar populations. 
In performing this analysis, another aspect to be investigated is the connection between
the  fundamental scaling relations mentioned above (the MZR and 
MSR) and galactic DS.   

We will also show the backward temporal evolution of the principal physical properties 
(i.e, gas fraction, metallicity, SFR, stellar mass) of star-forming galaxies that are part of the  local MZR and  MSR.
Other studies investigated the backward evolution of the MZR in star-forming galaxies with chemical
evolution models (\citealt{caluraMZ2009}, \citealt{lian2018ev}), but without focus on the interplay with other
scaling relations and on the particular role played by the MS. 

We will also analyze the temporal evolution and the fate of the galaxies with stellar masses, metallicities and SFRs
constrained by the  high redshift MZRs and MSRs. 
In particular, we will discuss the implications of the newly observed, flat MZR at redshift $z =2.2$ by \citet{mignoli2019},
who analyzed the metal content of 88 \civ-selected galaxies containing type 2 AGN with reliable measurements at high redshift.

Our paper is organized as follows.  In Section \ref{ES17}, we recall the main features of the chemical evolution model of  \citet{spitoni2017}. In Section \ref{meth}, the methodology used to create our grid of models is presented.
In Section \ref{obs} we present the main observational constraints considered in this work
and in Section \ref{results} we present our main results assuming a \citet{salpeter1955} IMF.  
Our conclusions are drawn
in Section \ref{conc}.

\section{The  model by  \citet{spitoni2017} } \label{ES17}

\citet{spitoni2017} presented  new analytical solutions in presence of an infall of gas
 that follows an exponential law for
the  metal abundances, gas and stellar masses within a galactic evolutionary framework
where instaneous mixing and IRA are assumed \citep{matteucci2012}. 
The infall of gas
that follows an exponential law is a fundamental assumption
adopted in most of the numerical chemical
evolution  models in which  IRA is relaxed.  Chemical evolution models
of our Galaxy (\citealt{caluraMZ2009, romano2010,grisoni2018,vincenzo2019, spitoni2016,spitoni2017GHZ,
spitoni2D2019,spitoni2019,spitoni2020}) assume that the various different stellar components
formed by means of separate accretion episodes of gas, with
the accretion rate of each episode expressed by an exponential law.

In \citet{spitoni2017}, the SFR is modelled by means of the \citet{schmidt1959} law, where the star formation rate $\psi(t)$ can be expressed as
$\psi(t)= S \cdot M_\mathrm{gas}(t)$,  
and where $M_\mathrm{gas}(t)$ is the gas mass at the time $t$ and $S$ is the so-called star formation efficiency (SFE),
which is commonly expressed in $\mathrm{Gyr}^{-1}$ and which is 
a free parameter of the model.

The gas infall rate  is expressed by an exponential law $\mathcal{I}(t) = A e^{-t/{\tau}}$,
where $\tau$ is the infall time-scale.
The quantity $A$ is a constant, constrained by the total infall gas mass $M_\mathrm{inf}$ \citep{spitoni2017}. 

In the model proposed by \citet{spitoni2017}, gas outflows in galaxies have been taken into account as well.
The outflow rate is  proportional to the SFR in the galaxy (see \citealt{recchi2008,spitoni2010,spitoni2015U}):
$W (t) = \lambda  \cdot \psi (t)$,
with  $\lambda$  being the loading factor parameter (a dimensionless quantity). 
 
\subsection{The analytical solution}
The analytical solution of \citet{spitoni2017} for the evolution of the gas-phase metallicity, defined as 
$Z = M_Z/M_\mathrm{gas}$,  is: 
\begin{eqnarray}\nonumber
Z(t)&=&  \frac{y_z  \, S\big( 1-R \big)}{ \alpha \tau-1}  \cdot\\
&& \frac{M_\mathrm{gas}(0) \, t \, \big( \alpha \tau - 1 \big)^2+ A \tau\big[t - \tau  (1  + \alpha  t)    +  \tau   e^{\alpha t -t/\tau}  \big]   }
{  A \tau  \big(e^{\alpha t -t/\tau} -1\big) + M_\mathrm{gas}(0)\big(
   \alpha \tau-1\big) }, 
\label{Znew}
 \end{eqnarray}
in which it is assumed that the infalling gas is metal-free, as is the
the galaxy at the epoch of its formation. 
In Eq. (\ref{Znew}) we introduce  the parameter $\alpha$, defined as $\alpha \equiv \big(1 + \lambda - R\big)S$.
The quantities $y_Z$ and $R$ are  the so-called yield per stellar generation and returned mass fraction
(for a detailed description, see \citealt{spitoni2017}), respectively.

The values of $y_Z$ and $R$  are average values computed at various metallicities and for various IMFs, and are
from Table 2 of \citet{vincenzo2016}, in which the compilation of stellar yields of \citet{romano2010} was considered. 
As for the stellar IMF, two different cases are studied in this work. 
In the case of a \citet{salpeter1955} we assume $R=0.287$  and  $y_O=0.018$,
whereas for a \citet{chabrier2003} (see Appendix \ref{section_chab}) $R=0.441$ and  $y_O = 0.0407$, where $y_O$ is
the oxygen yield per stellar generation. 

The other analytical expressions for the evolution of the total mass, gas mass and stellar mass
can be found in \citet{spitoni2017}.

More recently, \citet{weinbergO2017}  presented chemical analytical solutions adopting a different approach compared to  \citet{spitoni2017}, specifying  the  SFR history instead of an analytical form for the infalling gas.
The SFR derived by an 
exponential infalling gas rate in \citet{spitoni2017} is similar to the linear-exponential
SFR history  (proportional to  $t e^{-t/\tau_{\rm sfh}}$, where $\tau_{\rm sfh}$ is the time-scale of the exponential decline) 
presented by \citet{weinbergO2017}, and the results for the metallicity evolution are qualitatively similar.

\section{The methodology} \label{meth}

We focus on a given redshift, and we impose that the  galaxies must follow the observed MZR and
the star-forming MSR measured at that redshift 
(see Section \ref{obs} for details concerning the observational constraints).

First, we study the properties of the local galaxies by imposing the MZR and the MSR at redshift $z$ = 0.1,
and analyzing their `backward' evolution towards higher redshifts.
On the other hand, by imposing the same scaling relations at higher redshift, 
we will compute a `forward' evolution of the galaxies at lower redshifts and towards the present time.

In this framework, we will also study the evolution of the galaxies
which at redshift $z=2.2$ follow the flat MZR as derived by \citet{mignoli2019}. 

A grid of galactic models is created by varying some key parameters within a wide,
yet realistic space and with a fine resolution. 
The values adopted for the infall timescale $\tau$ span a range between 0.05 Gyr and 10 Gyr,
with a resolution $\Delta \tau$=0.05 Gyr as in \citet{spitoni2017}.
The  wind parameter  $\lambda$  can vary in the range between 0 and 10,
with a resolution  of $\Delta \lambda$=0.05, i. e. 10 times finer resolution than the one adopted in \citet{spitoni2017}, where  $\Delta \lambda$=0.5 .
Moreover,  as done in \citet{spitoni2017}  the total infall mass in the \citet{salpeter1955} case varies between 
$10^{7.5}$ and 10$^{11.5}$ M$_{\odot}$.  
In the case of the \citet{salpeter1955} IMF, we vary the infall masses accordingly to the following expression:
\begin{equation}
M_{inf}= (k/10) \cdot 10^{j+7.5} \mbox{  $M_{\odot}$}, 
    \end{equation}
where $1 \le k \le 100$, and steps $\Delta k=5$, $1\le j \le 3$ and steps $\Delta j=1$.
In the  Appendix \ref{section_chab}, we will discuss the effects of a different IMF on this parameter. 
Each galaxy has been evolved in time with a resolution of $\Delta t$=0.036 Gyr.

Following \citet{spitoni2017}, we do not assume that all galaxies are coeval, but we allow them to form at different redshifts.
At a fixed stellar mass, a galaxy is considered as part of the local (or high-redshift) MZR and MSR 
if the differences between the predicted  oxygen abundance and SFR values and the observed ones are  smaller than  $10^{-3}$: i.e.  $\Delta {\rm (O/H)}$=(O/H)$_{\rm model}-$  (O/H)$_{\rm data} < 10^{-3}$  dex  and $\Delta {\log ({\rm SFR )}}=\log ({\rm SFR}_{\rm model})-\log ({\rm SFR}_{\rm data}) < 10^{-3}$  $\log$(M$_{\odot}$ yr$^{-1}$).   In this way we are assuming null scatter in the MZR and MSR.

\begin{figure*}
\begin{centering}
\includegraphics[scale=0.38]{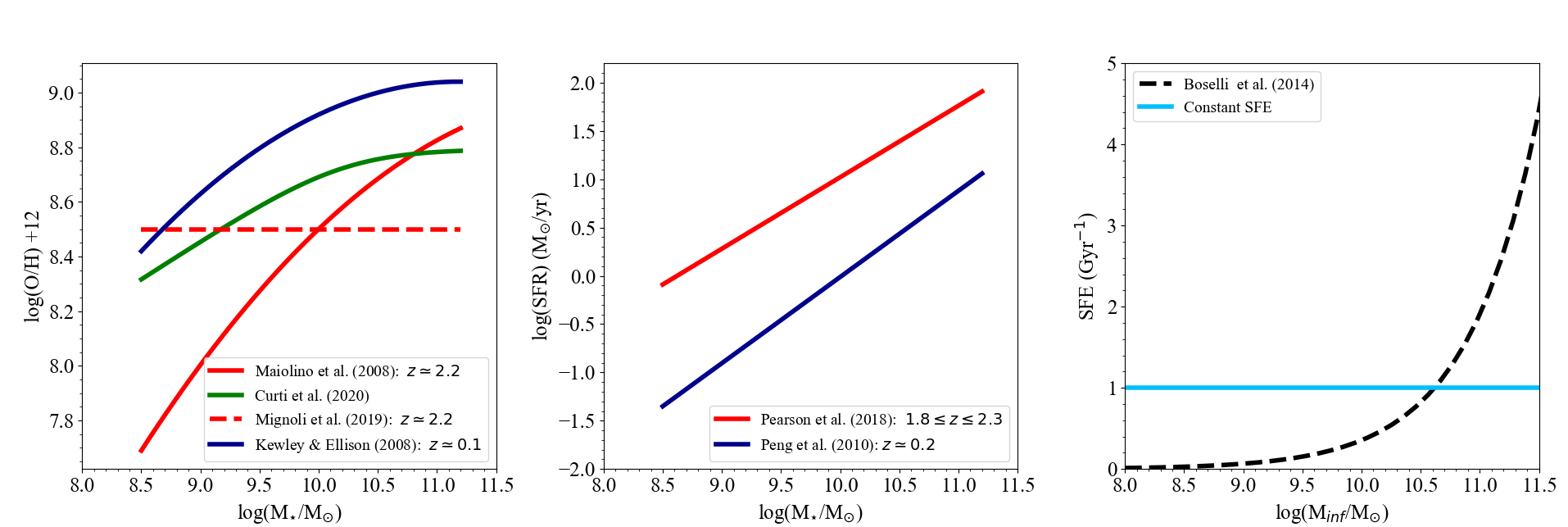}
\caption{{\it Left  panel:} Observed MZRs adopted in this paper at different redshifts: the local one by \citet{kewley2008} is depicted with the blue line, the high redshift MZR by \citet{maiolino2008} with the red one.
  The red dashed line is the MZR observed by \citet{mignoli2019} at $z\sim2$.
   With the solid green line we also show the local MZR relation for SDSS galaxies found by \citet{curti2020} with a fully T$_{\rm e}$-based abundance scale analysis.  {\it Middle panel:}  The star-forming main sequence by \citet{peng2010} in local SDSS galaxies is shown with the blue line, whereas the high redshift relation is the one by \citet{pearson2018}.
  {\it Right panel:} With the dashed black line we show the star formation efficiency as a function of the infall mass $M_{inf}$ from the scaling relation for the gas depletion timescale as found by \citet{boselli2014}.
  The constant SFE fixed at the value of 1 Gyr$^{-1}$ is indicated with the ligh blue curve.} 

\label{SFE1}
\end{centering}
\end{figure*}
Moreover, in order to be  consistent  with the observational data by \citet{trussler2020}, we define the average mass-weighted age following \citet{calura2014}.
For each galaxy, characterised at a certain evolutionary time $t$ by a SFR $\psi(t)$, 
we compute the average mass-weighted age at the time $t_n$  as: 
\begin{equation}
{\rm Age}(t_n) = \frac{\int_0^{t < t_n} (t_n-t) \psi(t) dt}{\int_0^{t < t_n} \psi(t) dt}.
\label{Age_mw} 
\end{equation} 

Average mass-weighted ages have been computed at four different redshift values,
each one corresponding to a different value for $t_n$. 
The star formation efficiency SFE is constrained by means of the scaling relation for local galaxies presented
by \citet{boselli2014} which link the typical galaxy gas depletion timescale, 
$\tau_\mathrm{gas}$,  and the galaxy stellar mass as follows: 
\begin{equation}
\log(\tau_\mathrm{gas})=-0.73\log( M_{\star}/ M_{\sun})+16.75,
\label{taugas}
\end{equation}
where $\tau_\mathrm{gas}=M_\mathrm{gas}/\mathrm{SFR}$ is defined as the inverse of our SFE, namely $\tau_\mathrm{gas} = 1/S$. 
According to Eq. (\ref{taugas}), galaxies with higher stellar mass would consume their available gas mass on shorter typical timescales 
if only star formation activity were taking place in the galaxy; this means that larger galaxies are expected to experience, on average, higher SFEs
(see \citealt{matteucci2012}). 

In our standard model, we adopt Eq. (\ref{taugas}) to constrain the galaxy SFE (which is kept fixed during the galaxy evolution), 
given an initial value for the galaxy infall mass, $M_\mathrm{inf}$
(in the right panel of  Fig. \ref{SFE1} the variation of the SFE as a function of the  total gas infall mass $M_\mathrm{inf}$). 
The role of this assumption will be discussed in Section \ref{SFE_const}.
We will also consider another case in which a constant SFE as a function of galaxy mass is assumed.

\section{Observed  scaling relations at different redshifts} \label{obs}

In this Section we describe the observational constraints for the MZR and the star forming MSRs at different redshifts adopted  in this work. 

The observed MZR at redshift  $z\sim 0.1$ is the one  by \citet{kewley2008}, as fitted  by \citet{maiolino2008}
 with a calibration  based on photoionization models provided by \citet{KD2002}:

\begin{equation}
\log({\rm O/H}) +12=-0.0864 \cdot\left[\log\left( \frac{M_{\star}}{M_{\odot}}\right) -\alpha \right]^{2}+\beta,
\label{fit}
\end{equation}
where $\alpha$ and $\beta$ are the free parameters of the fit that have been fixed at the values of $\alpha$=11.18 and  $\beta$=9.04. 

In Section \ref{sec_curti}, we will show how our results are sensitive to the adoption of a different calibration method
  for metallicity measures. 
In particular, we will show the effects of the adoption of a T$_{\rm e}$-based metallicity calibration for the local MZR.

As for the high redshift MZR, we will consider two different relations. 
The first is the one presented by \citet{maiolino2008} for galaxies at redshift $z= 2.2$, for which 
the analytical fit form is the same as Eq. (\ref{fit}) but with $\alpha=12.38$ and  $\beta=8.99$. 

 Concerning high redshift objects, \citet{maiolino2008} underlined that at that time no single strong line
calibration method existed over the wide metallicity range spanned by such  galaxies. 
Therefore,  at low metallicities (12 + log(O/H) < 8.35) they adopted calibrations
of strong-line diagnostics based on the T$_{\rm e}$ method, whereas 
at higher metallicity they relied  on photoionization models provided by \citet{KD2002}.

In our analysis, we will also consider the recent results by \citet{mignoli2019}, in which a flat MZR with log(O/H)+12 $\simeq$ 8.5 dex was
found for galaxies hosting Type 2 AGN at  $z \simeq 2.2$.

Finally, we present the scaling relations  of the star-forming main sequence that will be used as a further constraint of  our models.
At low redshift, the MSR is the one of  \citet{peng2010}, computed at redshift $z=0.2$,
\begin{equation}
\log({\rm SFR}/{\rm M}_{\odot} {\rm yr}^{-1}) = 0.89 \cdot\log\left( \frac{M_{\star}}{M_{\odot}}\right)-8.93.
\label{}
\end{equation}
The high redshift ($1.8 \leq z \leq 2.3 $)  star-forming MSR  is the one  of \citet{pearson2018}, fitted by the following expression:
\begin{equation}
\log({\rm SFR}/{\rm M}_{\odot} {\rm yr}^{-1}) = 0.74 \cdot\left[\log\left( \frac{M_{\star}}{M_{\odot}}\right) -10.5 \right]+1.39.
\label{}
\end{equation}
In  the left  and middle panels of Fig. \ref{SFE1} we show the observed  MZRs and the MSRs used in this work, respectively.

\section{Model results}\label{results}
Our aim is to reconstruct the past star formation history of the galaxies
which obey the two fundamental scaling relations considered here, i.e. the MZR and the MSR of star-forming galaxies,
at two representative different redshifts.
The redshifts at which we perform our analysis
are those which already include significant samples of 
star forming galaxies and for which the MZR has been  evaluated homogeneously, i. e. using the same 
method for both the stellar mass and the interstellar metallicity
as described in \citet{caluraMZ2009}.\\
The properties of the galaxies who follow the MZR and MSR at such representative redshifts 
will be studied
at different epochs:   the backward evolution of the galaxies which obey  the observed 
 MZR and MSR at $z\sim 0.1$ and the forward evolution of objects that at redshift  $z\sim 2.2$  are part of both the MSR and  the MZR.
Our aim is to track the evolution of the galaxies which, in a given phase of their history, follow at the same time the MZR and the MSR.

\begin{figure}
\begin{centering}
\includegraphics[scale=0.37]{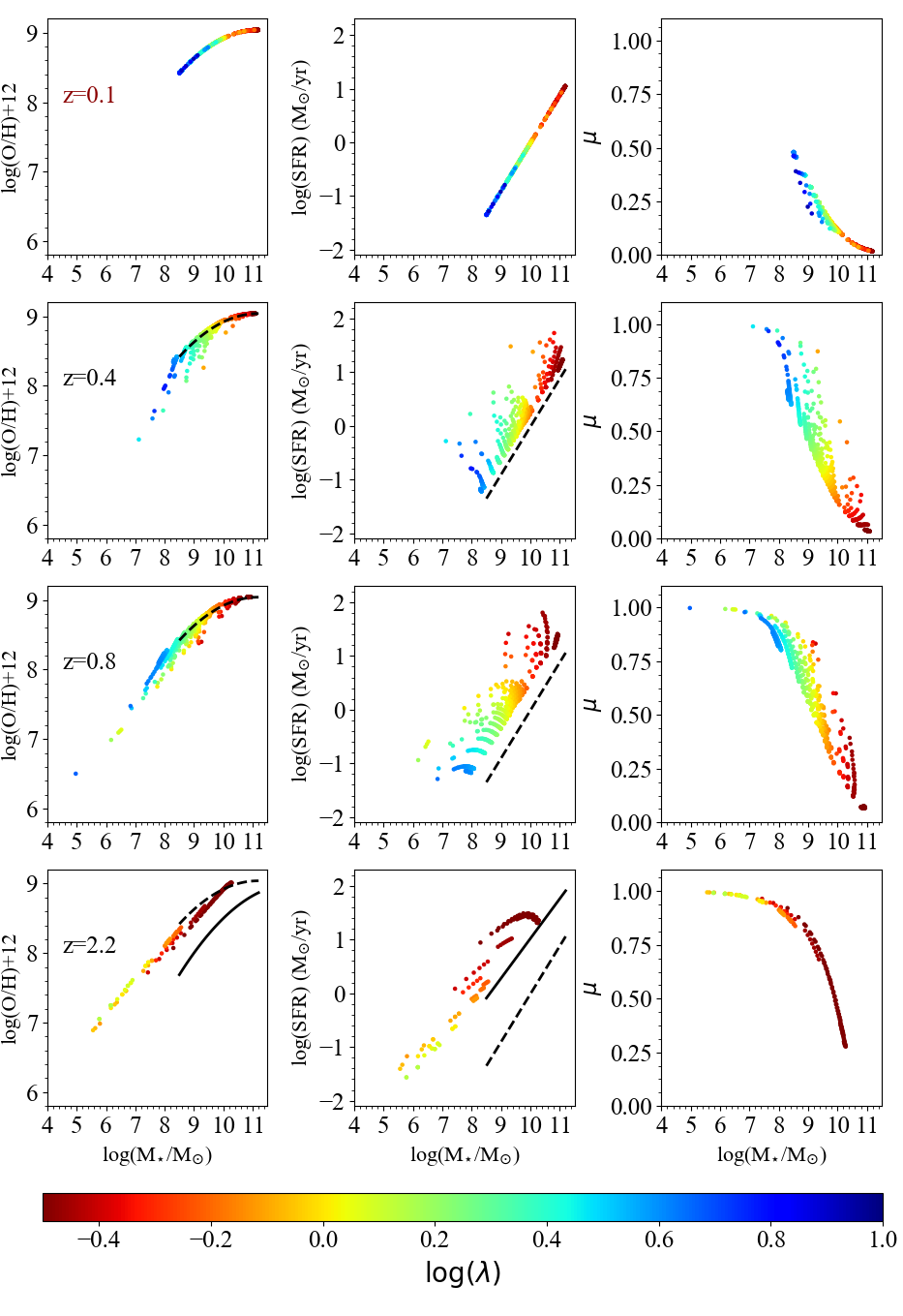}
\caption{Backward evolution from $z=0.1$ to $z=2.2$ of the galaxies which obey the analytical fit of the observed 
 MZR at $z\sim 0.1$ of \citet{kewley2008}, as reported by \citet{maiolino2008} and the local main sequence of star-forming galaxies as derived by \citet{peng2010}.
In the first, second and third column the evolution of MZR, SFR vs stellar mass and gas fraction vs stellar mass evolution are shown, respectively.  
The colour coding indicates the loading factor parameter $\lambda$.
 The black solid lines in the bottom left- and middle-panels indicate the MZR 
  and MSR observed at redshift $z=2.2$, respectively. The black dashed lines in the panels of the first column and in each SFR-log(M$_{\star}$/M$_{\odot}$) plot, show the MZR at redshift $z \sim 0.1$ of \citet{kewley2008} and the MSR derived in local star-forming galaxies by \citet{peng2010}, respectively.}
\label{z01_12p}
\end{centering}
\end{figure}

\begin{figure}
\begin{centering}
\includegraphics[scale=0.37]{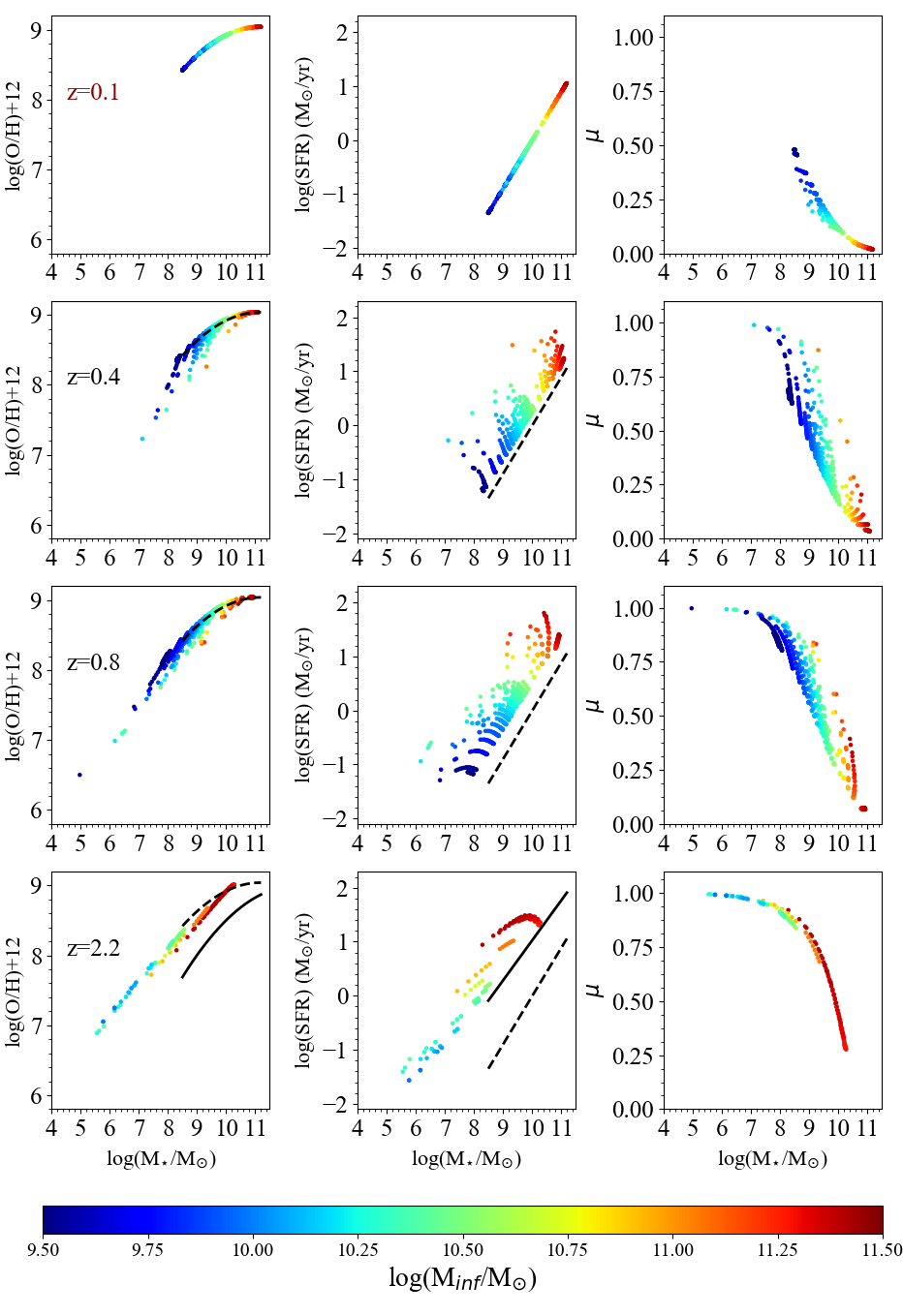}
\caption{As in Fig. \ref{z01_12p}, but the colour coding indicates the total infalling gas mass.}
\label{z01_12p_minf}
\end{centering}
\end{figure}

 \subsection{Backward evolution}

\subsubsection{The MZR and star-forming MSR at $z=0.1$}\label{MZ_01}\label{backward_sec}
Fig. ~\ref{z01_12p} and Fig. \ref{z01_12p_minf} show 
the backward evolution of the galaxies which obey the analytical fit of the observed 
MZR at $z\sim 0.1$ of \citet{kewley2008}, as reported by \citet{maiolino2008} and the main sequence of star-forming galaxies as derived by \citet{peng2010}.
In Fig. ~\ref{z01_12p} and in Fig. \ref{z01_12p_minf}, the galaxies are colour-coded
as a function of their loading factor $\lambda$ and infalling gas mass, respectively.

By imposing that they obey the \citet{peng2010} MSR and by means of the  \citet{schmidt1959} law which links the gas
mass to the SFR, we correctly retrieve the anticorrelation
between a fundamental quantity related to the gas accretion history of
galaxies, i.e. the gas fraction $\mu \equiv M_{gas}/(M_{gas}+M_{\star})$, and the stellar mass. 
It is important to note that such an anticorrelation is a well established,
empirical relation of galaxies found at various redshifts.
Locally, such anticorrelation was found by \citet{kanna2004}, 
who determined the gas fraction for 35,000 galaxies from the Sloan Digital
Sky Survey (SDSS) and the Two Micron All Sky Survey
(2MASS) databases on the basis of photometric techniques. 
A similar study on a smaller sample of low-mass galaxies has been
performed by \citet{geha2006}, confirming the results of \citet{kanna2004}. 
The slope of the curve steepens from low-mass to high-mass systems,
which is generally interpreted with 
decreasing gas consumption timescales from dwarf to giant galaxies \citep{calura2008}.\\
The past evolution of our model galaxies has been computed at three different redshifts, i.e. $z=0.4$, 0.8 and 2.2.\\
The larger redshift value present in the plot is the one at which a substantial amount of galaxies have appeared,
which now are star-forming and build the local MZR. The epoch at which each galaxy must be born 
is a result of our method, and which stems directly from Eq.~\ref{Znew} and from the imposition of
the observational constraints.\\
A significant evolution of all the three quantities shown in Fig. ~\ref{z01_12p} is evident from the plot.
A positive correlation between stellar mass and metallicity is already in place at $z=2.2$.
In the young galaxies present at that redshift, the shape of the MZR is missing the characteristic plateau
at large masses which is generally observed at all redshifts.

 This can be understood considering that the one at $z=2.2$ as shown in Fig. ~\ref{z01_12p} is only a sub-sample of the entire population of all the star-forming
galaxies at that redshift, which will obey the MZR but which will include also 
many systems which now are passive, i.e. which do not belong to the local MSR relation,
and which build the high-mass plateau at  that redshift,
as shown in \citet{savaglio2005} and \citet{maiolino2008}.

The relation between SFR and mass appears appreciably flatter than at $z=0.1$,
and even if the gas fraction still anti-correlates with the stellar mass, it is characterised by 
an inverted curvature with respect to the local relation. All the galaxies building the three relations are 
characterised by very low wind efficiencies ($\log(\lambda)\le -0.4$) and by the large infall mass values
($\log(M_{inf}/M_{\odot})\sim 11.5$). 
 \begin{figure}
\begin{centering}
\includegraphics[scale=0.43]{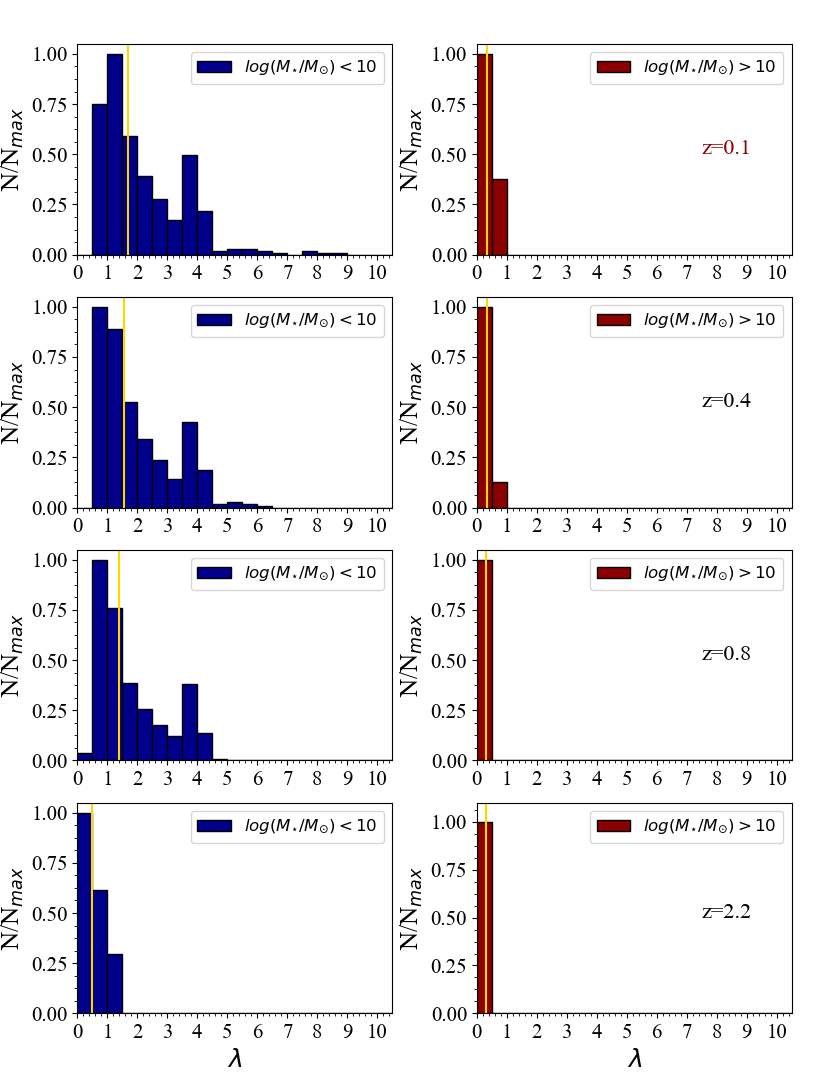}
\caption{Distribution of the loading factor parameter $\lambda$ for the galaxies of Fig. \ref{z01_12p} 
  computed at redshifts $z=2.2$, $z=0.8$, $z=0.4$, and $z= 0.1$ for 2 different bins of stellar mass.
  The left panels show distributions for objects with stellar mass $log(M_{\star}/M_{\odot})< 10$, and  right panels indicate distributions for stellar mass  $log(M_{\star}/M_{\odot})> 10$.
  The yellow vertical lines indicate the median values of each distribution.}
\label{D_lambda_masses}
\end{centering}
\end{figure}

At $z=0.8$, the MZR appears more extended than at $z=2.2$, in particular in that it is much more populated 
at $\log(M_{\star}/M_{\odot})\ge 8$. 
At this redshift, the hint of 
a plateau has already appeared in the MZR at the highest mass values, namely at $\log(M_{\star}/M_{\odot})> 10$.
The appearance of the plateau may be interpreted with a flattening of the star formation history
of the most massive galaxies, which are characterised by the largest SFR values 
and shortest gas consumption timescales (see Fig. \ref{10_5_time}). 
As a consequence of a faster gas consumption 
in the largest galaxies, their present-day metallicity is reached at earlier times than in low-mass galaxies.

Also the SFR-$M_{*}$ relation of the galaxies at $z=0.8$ is more extended and more scattered than at $z=2.2$. \\
The $\mu-M_{\star}$ relation sees the appearance of a significant population of newly born, extremely gas-rich and metal-poor galaxies characterised
by $\mu\sim 1$ at $\log(M_*/M_{\odot})< 8$. 
Also the range of the wind parameter and infall mass values presented by the galaxies at $z=0.8$ have extended 
considerably with respect to $z=2.2$. 
The maximally gas-rich galaxies present the largest 
values of the $\lambda$ parameter ($\log\lambda \sim 0.6$) and are also characterised by the lowest infall mass values.\\
On the other hand, the strong evolution experienced by the largest systems leads them to keep their gas fractions very low.  \\
At $z=0.4$ the horizontal extension of the the MZR has reduced, as all galaxies have grown in stellar mass. 
The MZR is now much more similar to the local one, with the exception of
a larger scatter at low masses which, by construction, later on will disappear. 
The population of galaxies with gas fractions $\sim 1$ have considerably reduced.
The gas fraction-mass relation shows a parabolic behaviour at the largest stellar masses,
with an inflection at $\log(M_{*}/M_{\odot}) \sim 9$.

\begin{figure}
\begin{centering}
\includegraphics[scale=0.43]{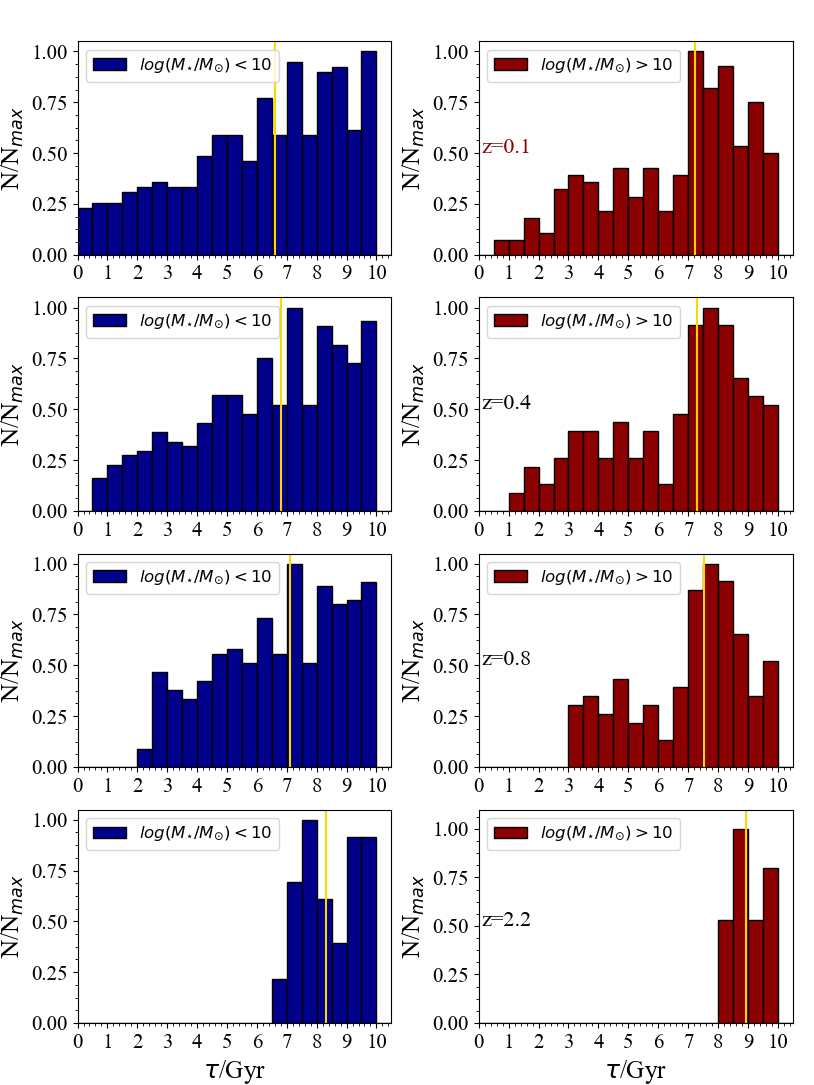}
\caption{As in Fig. \ref{D_lambda_masses}, but in each panel the plotted distribution is for the infall time-scale parameter $\tau$. }
\label{D_tau_masses}
\end{centering}
\end{figure}

In Figs. \ref{D_lambda_masses} and \ref{D_tau_masses}, we present 
the distribution of the loading factor parameter $\lambda$ and the infall time-scale parameter $\tau$
at different redshifts for the model galaxies of Figs. \ref{z01_12p} and \ref{z01_12p_minf}.  
The distribution are computed in two different bins of stellar masses
($M_{\star}<10^{10} M_{\odot}$ and $M_{\star}>10^{10} M_{\odot}$). \\
The  galaxies present at redshift $z=2.2$ are characterized by  weak winds and long time scales of
gas accretion. 
In fact, at low redshift these objects have to  maintain
a large reservoir of gas in order to be part of the local MSR.

Galaxies in the larger mass bin show very weak winds and very little evolution of both the median $\lambda$ and $\tau$ values.

\begin{figure}
\begin{centering}
\includegraphics[scale=0.45]{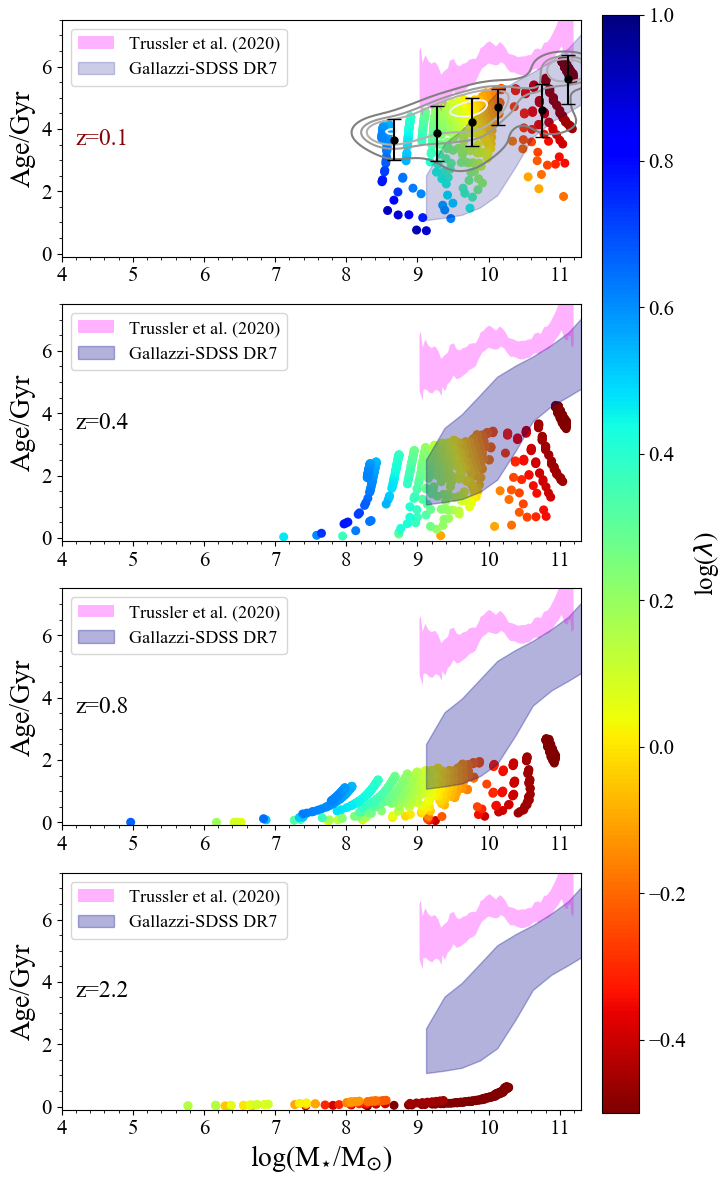}
\caption{Backward evolution of the mass-weighted age versus stellar mass relation (computed at $z= 2.2$, $z= 0.8$, and $z= 0.4$) for the galaxies
which lie on the MZR  of \citet{kewley2008} and on the MSR as derived by \citet{peng2010}. 
The colour coding
indicates the  loading factor parameter $\lambda$. The white-grey contour lines indicate isodensity contours. 
 The shaded pink area indicates the observational data by \citet{trussler2020} for local star-forming galaxies ($0.02<z<0.085$), whereas the shaded blue area stands for
the relation obtained using the SDSS-DR7 catalog of mass-weighted ages estimated as in \citet{gallazzi2008} using the same selection of star-forming galaxies as \citet{trussler2020}.
In the upper panel, the black points  are the mean age values  of the simulated galaxies at redshift $z \sim 0.1$ inside  bins of size   0.5 log(M$_{\star}$/M$_{\odot}$) and the error bars are the standard deviations. }
\label{trus_sal}
\end{centering}
\end{figure}

\begin{figure}
\begin{centering}
\includegraphics[scale=0.35]{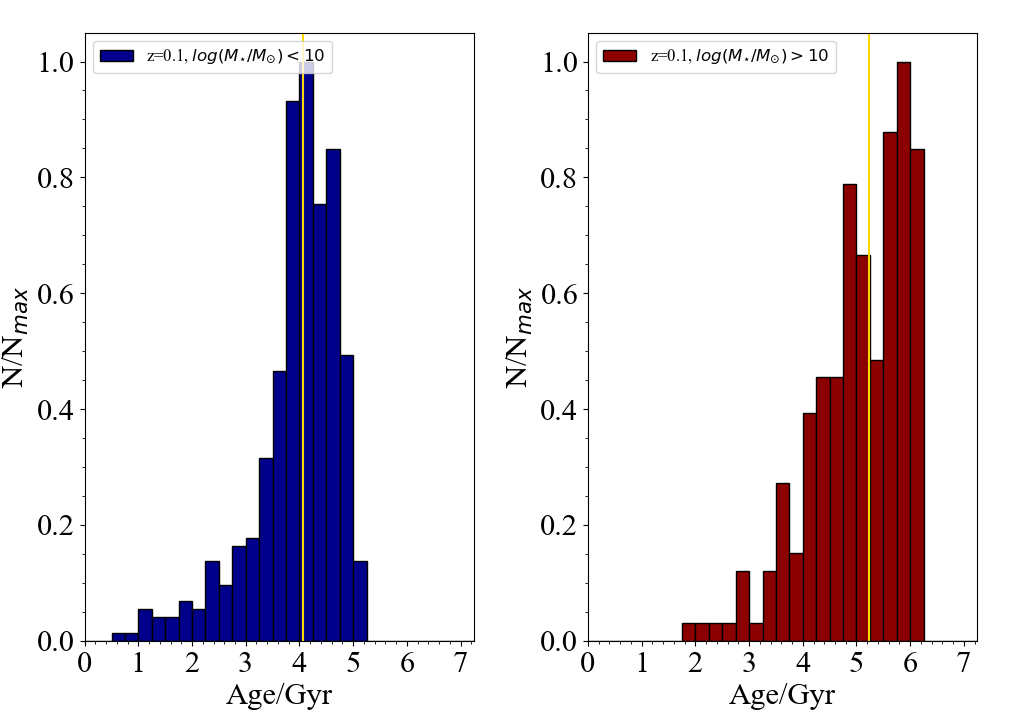}
\caption{Age distribution computed at redshift $z=0.1$ for the galaxies
which lie on the MZR  of \citet{kewley2008} and on the MSR  as derived by \citet{peng2010} in two different mass bins, $log(M_{\star}/M_{\odot})< 10$ (left) and
$log(M_{\star}/M_{\odot})\ge 10$ (right). 
In each panel, the yellow vertical line indicates the median value of the distribution.}
\label{age_D}
\end{centering}
\end{figure}

We note that galaxies characterized by  $M_{\star}<10^{10} M_{\odot}$ at redshifts $z<2.2$ maintain their stellar mass below this
value during their entire evolution up  to redshift $z=0.1$, as visible in Fig. \ref{z01_12p} and \ref{z01_12p_minf}.

\begin{figure*}
\begin{centering}
\includegraphics[scale=0.45]{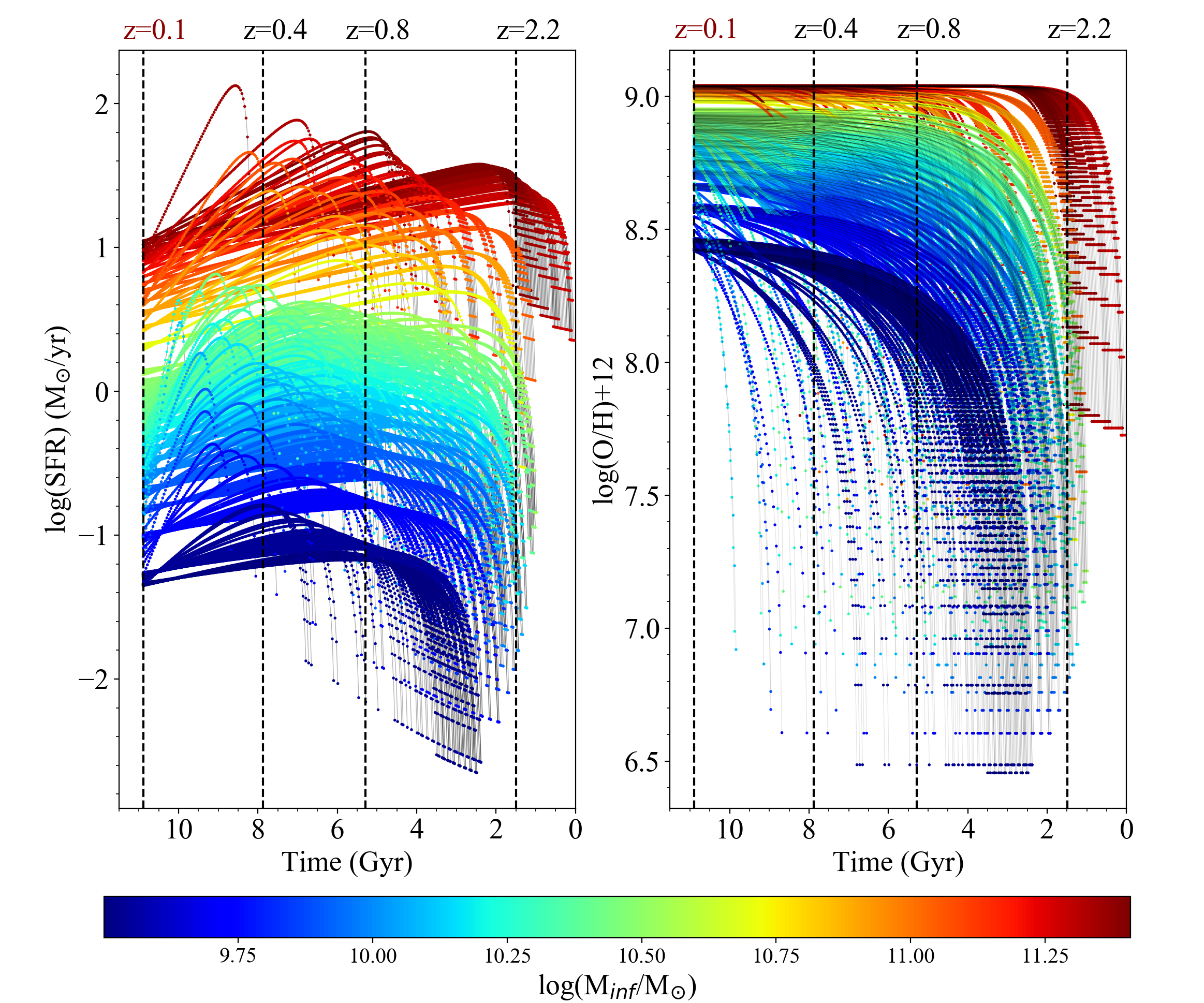}
\caption{Evolution of star formation rate (left panel) and oxygen abundance (right panel)  as a function of time  for the galaxies
which lie on the MZR of \citet{kewley2008} and on the MSR  as derived by \citet{peng2010}. The colour coding represents  the infall mass.}
\label{SFH_O_t}
\end{centering}
\end{figure*}
\begin{figure*}
\begin{centering}
\includegraphics[scale=0.4]{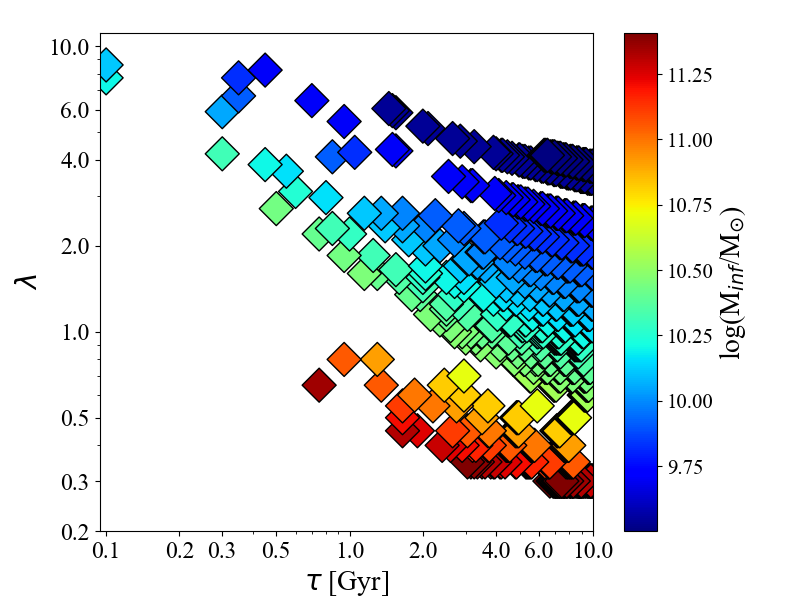}
\includegraphics[scale=0.4]{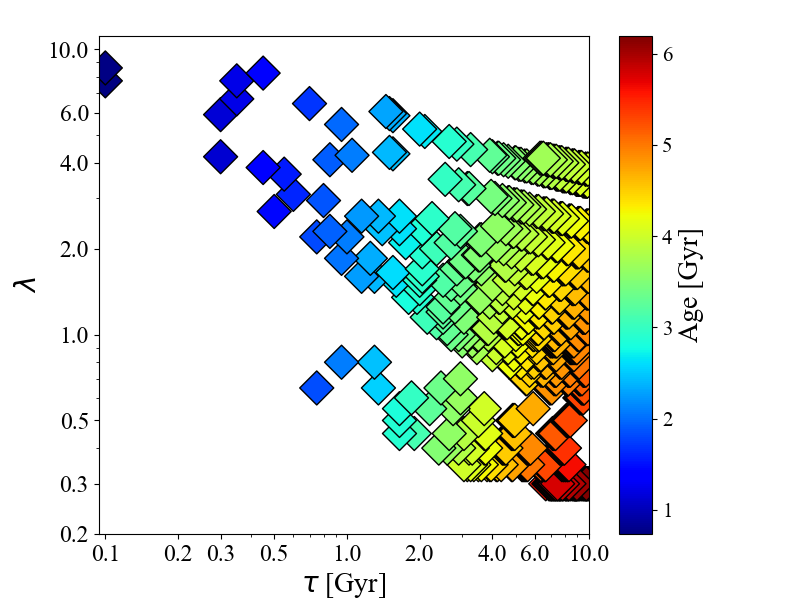}
\caption{Parameter space analysis (wind parameter $\lambda$ versus infall time scale $\tau)$  of the galaxies which line on the  
MZR of \citet{kewley2008} and on the main sequence of star-forming galaxies as derived by \citet{peng2010}. In the left panel the colour coding represents the infall mass, whereas in the right one it represents the galactic age.  }
\label{space_par}
\end{centering}
\end{figure*}

\begin{figure*}
\begin{centering}
\includegraphics[scale=0.5]{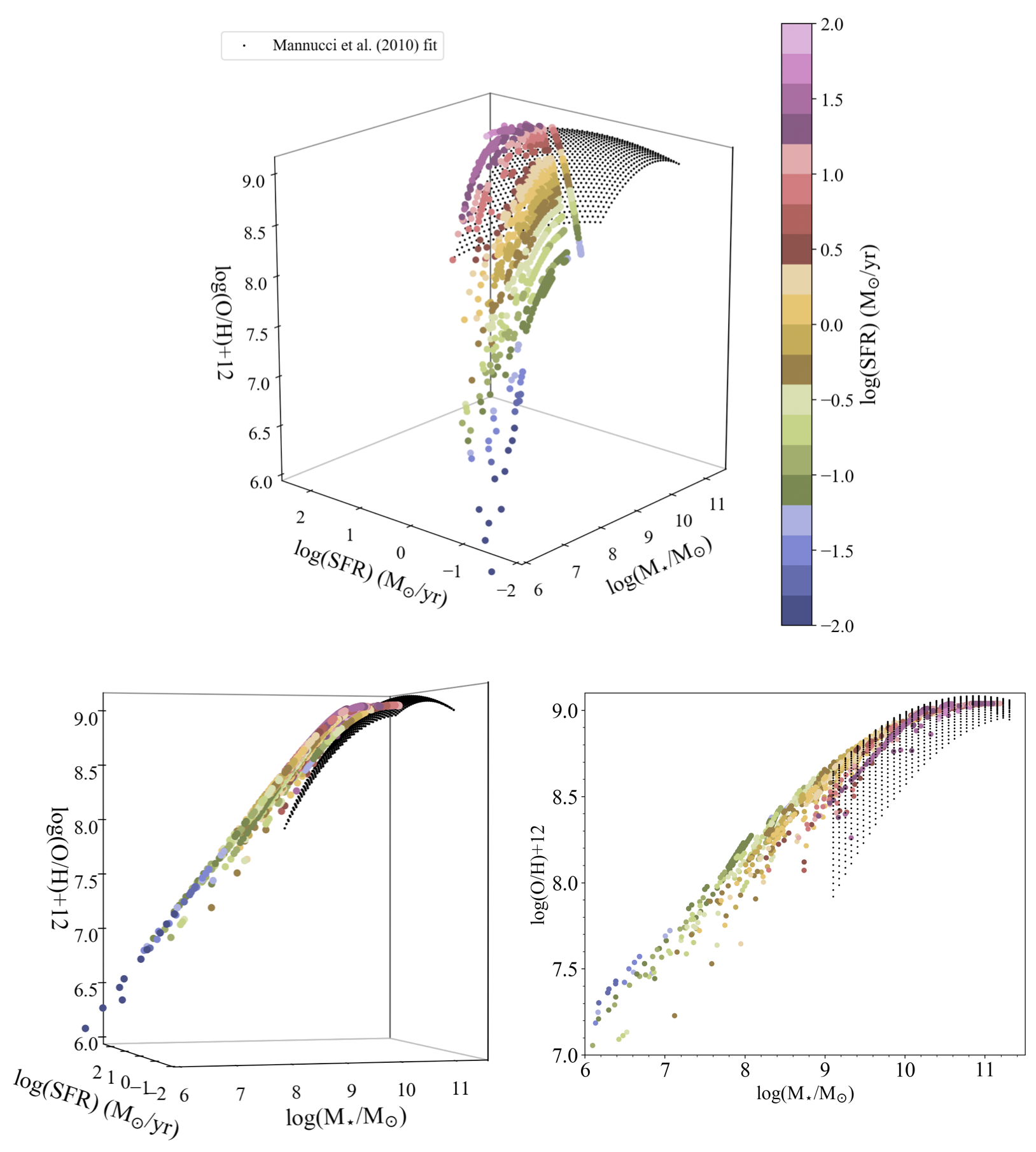}
\caption{ Three projections of the Fundamental Metallicity Relation of  \citet{mannucci2010} 
compared with the properties of the ancestors of the galaxies that at z=0.1 follow the MSR and MZRs. 
The big solid circles are the model galaxies illustrated in Section \ref{MZ_01}, and for which the SFR, stellar mass and metallicity values
reported in the plots are computed at z=0.1, 0.4, 0.8, 1.5 and 2.2, 
colour-coded as a function of their SFR values.  
The small black dots are a second-order fit of the SDSS data as presented in \citet{mannucci2010}. 
} 
\label{FMR_SFR}
\end{centering}
\end{figure*}

\begin{figure*}
\begin{centering}
\includegraphics[scale=0.35]{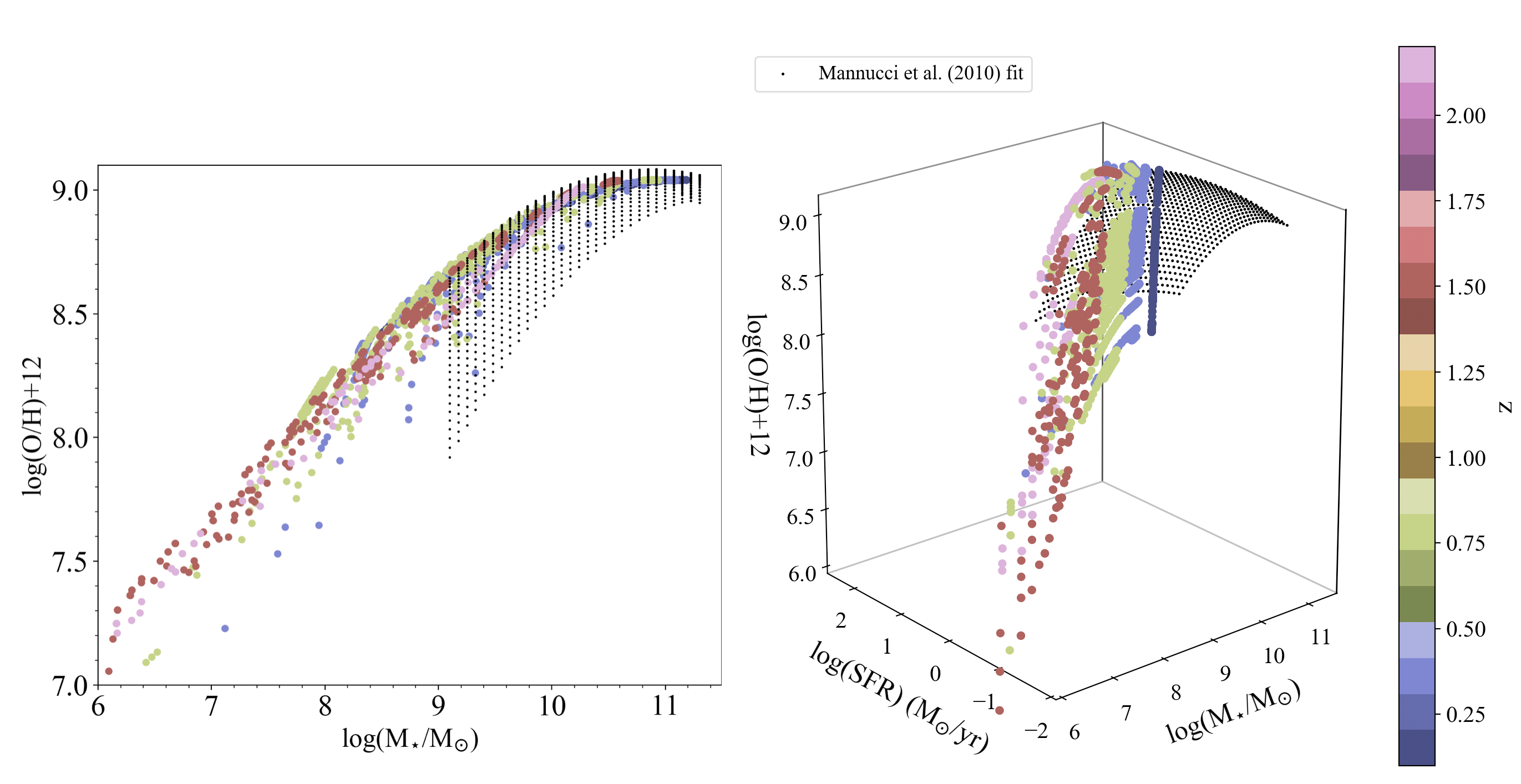}
\caption{ Two projections of the Fundamental Metallicity Relation of \citet{mannucci2010} compared with the properties of the model
    galaxies as in Fig. \ref{FMR_SFR}, colour-coded as a function of their redshift $z$. } 
\label{FMR_Z01}
\end{centering}
\end{figure*}

\begin{figure}
\begin{centering}
\includegraphics[scale=0.47]{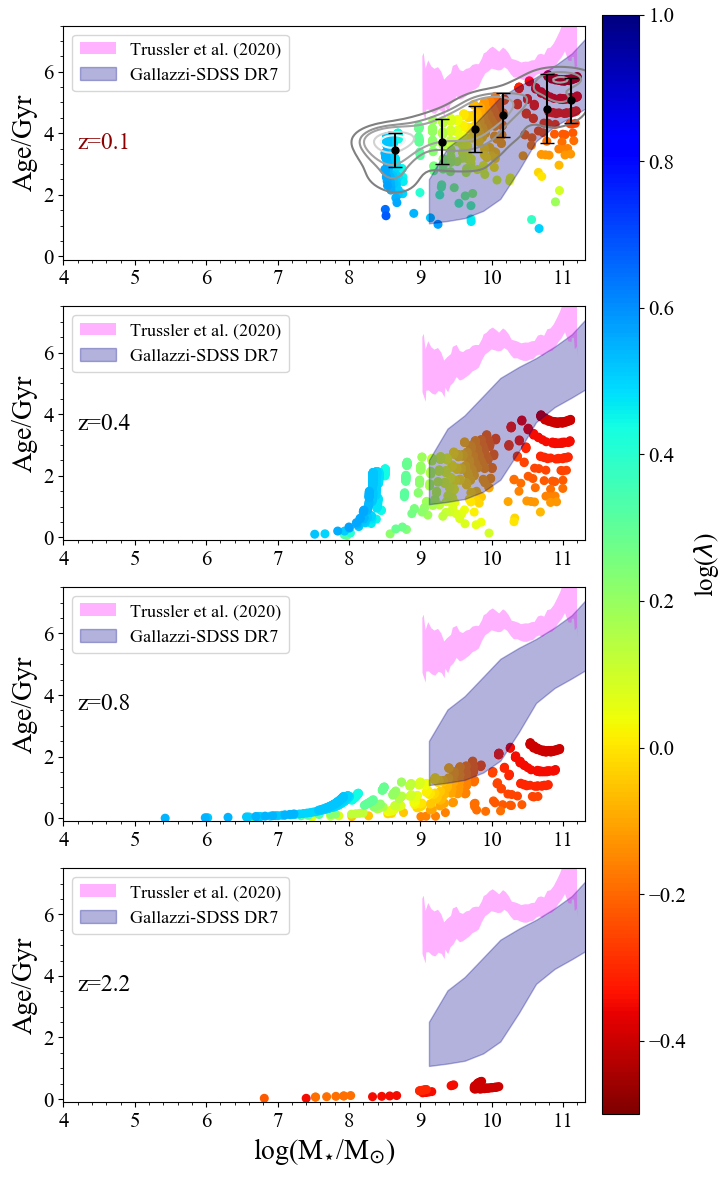}
\caption{As in Fig. \ref{trus_sal} but computed assuming a constant SFE fixed at the value of 1 Gyr$^{-1}$.}
\label{trus_sal_S1}
\end{centering}
\end{figure}

\begin{figure}
\begin{centering}
\includegraphics[scale=0.4]{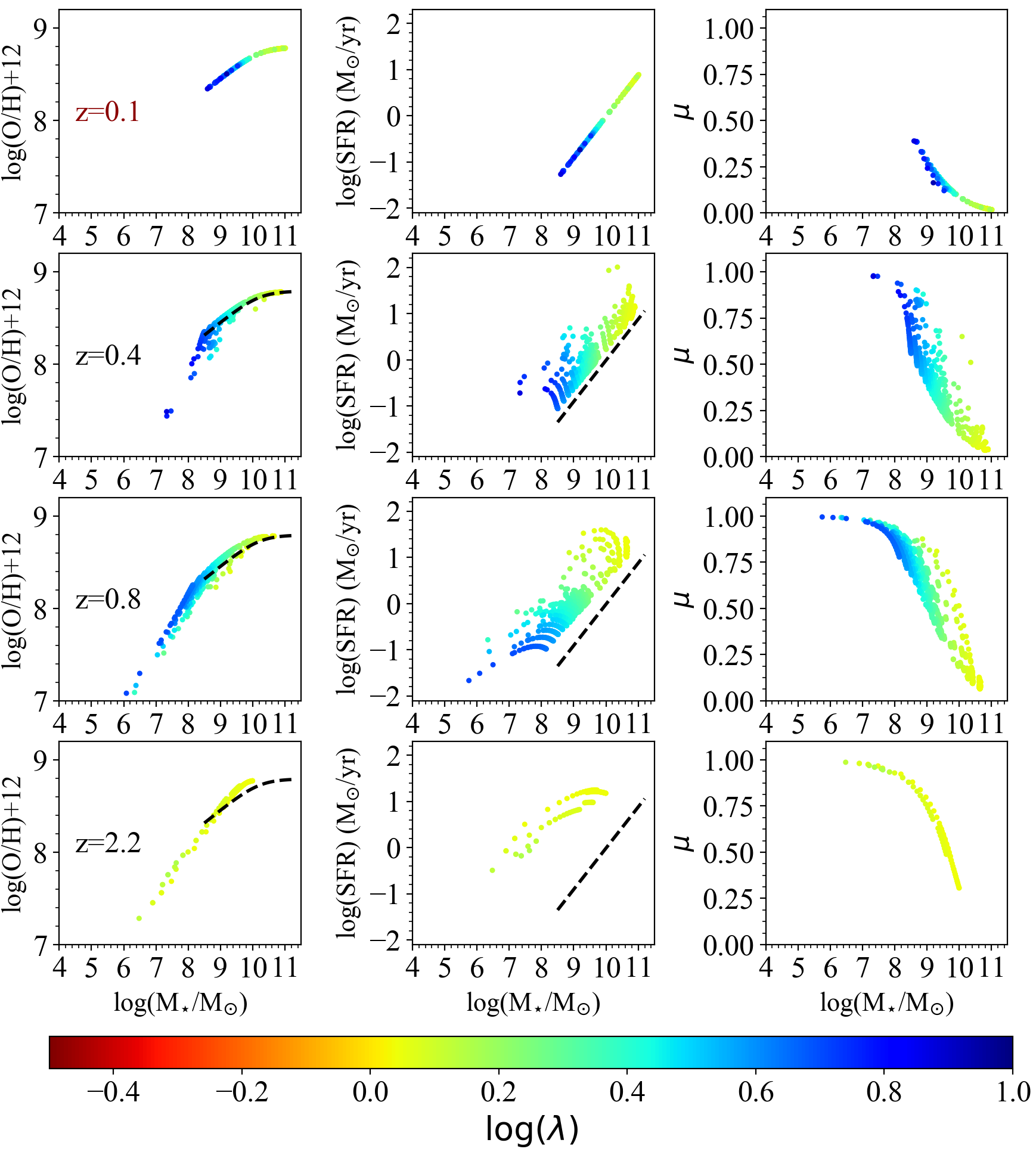}
\caption{
Backward evolution from $z=0.1$ to $z=2.2$ of the galaxies which obey the analytical fit of the observed 
MZR at $z\sim 0.1$ of  \citet{curti2020} and the local main sequence of star-forming galaxies as derived by \citet{peng2010},
colour coded as a function of the loading factor parameter $\lambda$
as in Fig.~\ref{z01_12p}. 
The black dashed lines in the panels of the first column and in each SFR-log(M$_{\star}$/M$_{\odot}$) plot, show the MZR of \citet{curti2020} and the MSR derived in local star-forming galaxies by \citet{peng2010}, respectively.}
\label{curti_34}
\end{centering}
\end{figure}

\begin{figure}
\begin{centering}
\includegraphics[scale=0.4]{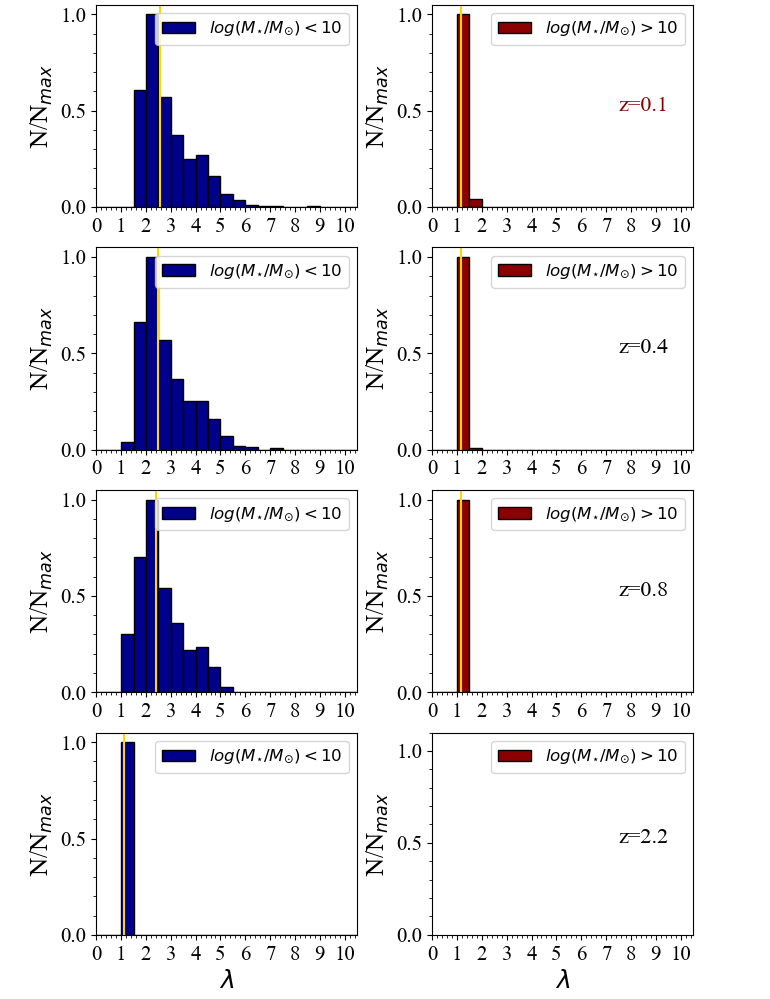}
\caption{ As in Fig. \ref{D_lambda_masses}, but computed with the local MZR proposed by \citet{curti2020}.}
\label{curti_lambda}
\end{centering}
\end{figure}

From Fig. \ref{D_tau_masses} we see that the low edge  
and the median of the infall time-scale distribution decrease with decreasing redshift in both stellar mass bins. This reflects that at each redshift,
 the galaxies - among the ones that are on the local MZR and MS -  with short time-scales of accretion $\tau$ and/or high loading factors did not exist yet at high z.

In the lowest mass bin, a decrease of the median $\tau$ is accompanied by an increase of
the median $\lambda$ value. This reflects the fact that a larger amount of mass with respect to more massive systems needs to be expelled 
to keep their metallicity at low values and to still have them on the MZR.

The median values of the infall time scale $\tau$  at  redshift $z=0.1$ are 6.59 Gyr  for galaxies with $M_{\star}<10^{10} M_{\odot}$
and 7.25 Gyr for  $M_{\star}>10^{10} M_{\odot}$, respectively (see Fig. \ref{D_tau_masses}). Such values are larger 
than the ones found in star forming galaxies by \citet{spitoni2017}, in which 75 \% of the galaxies showed infall time scales smaller than 6 Gyr. 
The difference is due to the fact that here we impose that the galaxies have to follow simultaneously the MZR and the MSR,
whereas in \citet{spitoni2017} the only requirement beside the MZR was a specific SFR value above a given threshold, which was $2.29 \times 10^{-11} {\rm yr}^{-1}$.

More massive galaxies on the MSR have larger SFRs, and as already mentioned, only these systems can afford the longest 
time-scales of accretion, which allow them to exhibit large SFR values down to $z=0.1$.

\subsubsection{Downsizing and scaling relations} \label{DS}

Fig. \ref{trus_sal}  shows the backward evolution of the age versus stellar mass relation for galaxies which at $z\sim 0.1$ obey the 
MZR and the main sequence of star-forming galaxies.

Approximately,  89.2 \% of the simulated galaxies are located in the region enclosed by  the most external contour line in the upper panel of Fig. \ref{trus_sal}. 
Such a positive relation  between age and  stellar  mass  is retained also if we increase the resolution of our parameter grid $(\lambda, \tau)$ by a factor of two. 
 We notice that the positive correlation between age and stellar mass is preserved at any redshift.

From the age distributions computed at redshift $z=0.1$ for the two range of stellar masses (see Fig. \ref{age_D})
we infer that the predicted median age of star-forming systems are $\sim$ 4.06 Gyr (lower stellar mass bin) and 5.23 Gyr (high stellar mass bin).

 The simulated galaxies at redshift $z=0.1$ reproduce the slope of the observed
age-mass relation for star-forming galaxies as found by \citet{trussler2020}. 
However, the predicted ages are systematically $\sim$ 1 Gyr younger    than the \citet{trussler2020} ones.

 We also compare the predicted age-mass relation with the one obtained adopting the mass-weighted ages estimated as in \citet{gallazzi2008} for SDSS-DR7 (Gallazzi et al. 2020, in prep - see also \citealt{pasquali2019}). From the Gallazzi-DR7 catalog we select star-forming galaxies with the same criterion as \citet{trussler2020} and over their same redshift interval. The Gallazzi relation is steeper and overall shifted to lower mass-weighted ages than the estimates by \citet{trussler2020}.

The noticeable differences in both normalization and slope of the Gallazzi age-mass relation and the \citet{trussler2020} relation can originate from different assumptions in the observational estimates of ages: i) the \citet{BC03} SSP models in comparison to \citet{maraston2011}, ii) the use of a stochastic library of parametric SFHs compared to a linear combination of SSP without assuming a functional form for the SFH, translating into differences in the allowed fraction of old stars, iii) limiting the formation age of the models to the age of the Universe compared to allowing SSPs as old as 15 Gyr, iv) a Bayesian statistical approach compared to a $\chi
^2$-minimization code.
While it is clear that a quantitative description of the mass-weighted age relation with mass is subject to the observational method adopted, both determinations agree in indicating a positive correlation.

Fig. \ref{trus_sal} implies a key result, i. e. that galactic downsizing as reflected from
the ages of their stellar populations is naturally accounted for by systems that obey two fundamental
scaling relations, namely the MZR and the MSR. 
 It is  worth noticing that assuming two scaling relations with zero scatter,  those together predict an age-mass relation with intrinsic scatter ($\sim$ 0.8 Gyr). This means that there might be a variety of SFHs that can lead to the current equilibrium between SFR and metal enrichment.
\begin{figure}
\begin{centering}
\includegraphics[scale=0.45]{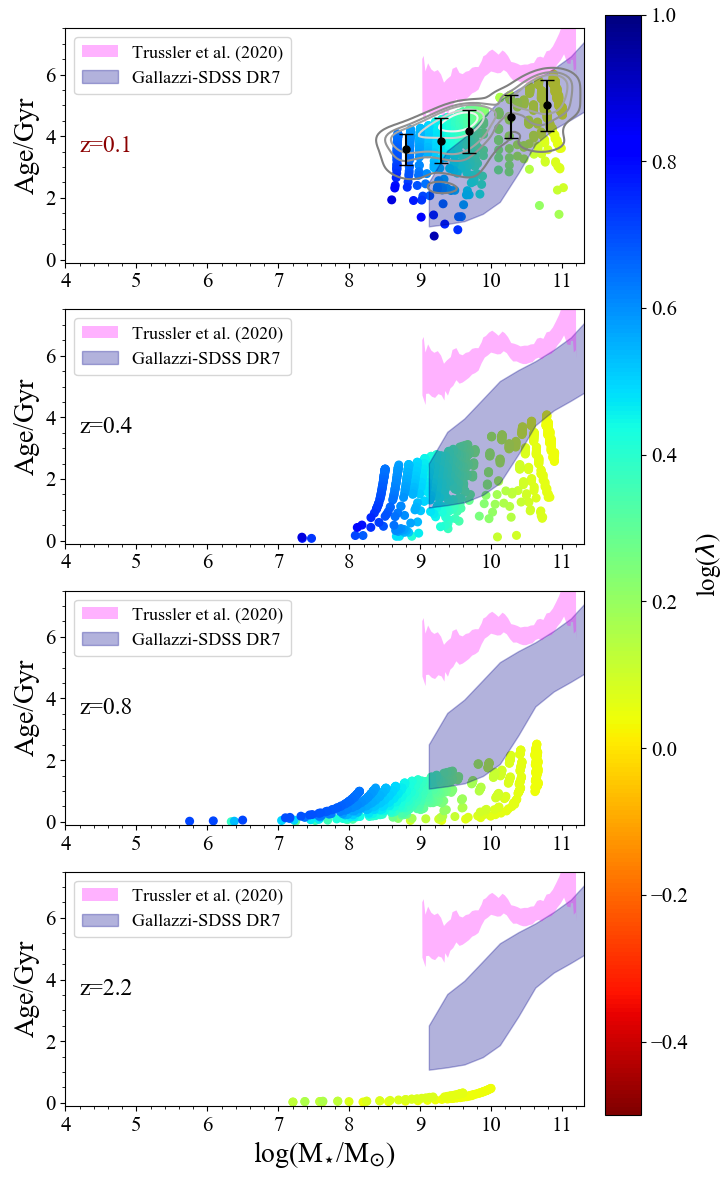}
\caption{ As in Fig. \ref{trus_sal}, but computed assuming at redshift $z \sim 0.1$ the  MZR proposed by  \citet{curti2020}.} 
\label{curti_trus}
\end{centering}
\end{figure}

The DS of galaxies is also visible from  
the temporal evolution of the SFR and oxygen abundances of the model galaxies as shown in Fig. \ref{SFH_O_t}.
In this figure, the evolution of each quantity is shown for systems of different masses as a function of time, for an evolutionary interval of $\sim~10.5$ Gyr.

These plots show that the bulk of the stars of the galaxies belonging to the highest mass end ($\log(M_{inf}/M_{\odot})>11$) of
the MZR must have formed at early times. In most of these systems the SFH peaks soon after their formation and
their metallicity saturates at its maximum value ($12+\log({\rm O/H})\sim 9$) 
after $\sim 1$ Gyr of evolution. Some high-mass galaxies may be characterised by SFHs which peak 
at more recent epochs (i.e. at ages $<4$ Gyr), and present SFR values larger than the ones of the older galaxies. 
The systems also have short infall timescales and populate the 
lowest $\tau$- ($<6$ Gyr) tail of the infall timescale distribution.
Moreover, they present $\lambda$ values larger than their older analogues but still $<1$ (see later), implying that galactic outflows are not the main driving mechanisms of their evolution.

From Fig. \ref{SFH_O_t}, we see that the lower the stellar mass of the galaxies, 
the lower the age at which the SFH peaks and the slower the growth of the metallicity with time.    
Our analysis of the star formation history of these galaxies  
is in agreement with the one presented in other works 
\citep{maiolino2008,caluraMZ2009} based on studies of the MZR, although the MZR alone does not necessarily imply galactic downsizing. 
In order to account for it one must impose simultaneously the MZR and the MSR. 
This represents one major result of the present work. 
With our models, it is also easy to show how the positive correlation between age and mass presented in Fig. \ref{trus_sal}
disappears completely if we
impose that galaxies must follow the MZR only.

As mentioned in Section \ref{meth},  in our model  the temporal evolution
of the  SFH can be well approximated with a function proportional to
$t e^{-t/\tau}$ (see discussion in \citealt{weinbergO2017}).  Hence, adopting this analytical expression,  the
mass weighted age presented in Eq. (\ref{Age_mw})  computed at the
time $t_n$  can be rewritten as follow:
\begin{equation}
 {\rm Age_{\rm approx}}(t_n) = \frac{t_n}{1-e^{-t_n/\tau}}-\tau.
\label{Age_analitical}   
 \end{equation}

In Fig. \ref{SFH_O_t}, we note that the  oldest galaxies have been
 evolving for  roughly  $t_n \simeq 10.5$ Gyr. Moreover, these objects
are characterized by long time-scale of accretion $\tau$, and a
representative value  is  $\tau \simeq 9.5$ Gyr (see
Fig. \ref{D_tau_masses}). Substituting these quantities in the
approximated  mass-weighted age introduced above
 in Eq. (\ref{Age_analitical}), we find a value of  ${\rm Age_{\rm
    approx}}(t_n) = 6.19$ Gyr, in perfect agreement   with our results
for the oldest objects as reported in Fig. \ref{trus_sal}.

To better understand the main reason for the results discussed here, 
in Fig. \ref{space_par} we show the loci occupied by our galaxies in 
the space of the free parameters of our model, i. e. $\lambda$ and $\tau$. 
The most massive objects are the oldest ones, which are generally located in the bottom-right corner of the plots and which are 
characterized by longest  accretion timescales and  smallest values for the wind parameter.
A population of young, very massive systems is visible in Fig. \ref{space_par}, which are
the galaxies with the largest SFR values of Fig.  \ref{SFH_O_t}. 
In Fig. \ref{space_par}, such systems are considerably underdense with respect to their older analogues.
This occurs because, in order to follow both the MZR and MSR, only a very limited range of values in the $\lambda$-$\tau$ space will be allowed.
In this particular region of this space, small perturbations of the values of each of these parameters will be enough
to drive galaxies off one (or both) of the two fundamental scaling relations, which explains the paucity of points with $\tau \sim 1$ Gyr and
$\lambda \sim 0.8$. On the other hand, a larger range and larger variations for each parameter will be admitted in the
bottom-right corner, where massive system tend to cluster. 

In galaxy evolution models, some properties which reflect galactic DS can generally be accounted for in different 
ways and acting on various parameters, i. e. by varying the stellar IMF (\citealt{calura2009}), or 
by varying the SF efficiency in galaxies of various mass \citep{matteucci1994}. 
In the next section we will discuss further the role of this parameter in particular, as well as the implications on 
our results.

\subsubsection{The Fundamental metallicity relation of \citet{mannucci2010}}\label{sec_FMR}

 \citet{mannucci2010} showed that star forming galaxies lie on a tight surface in a 3D space defined by stellar mass, gas-phase metallicity, and SFR.
The median values of metallicity of the SDSS galaxies can be expressed by means of 
a second-order polynomial fit in M$_{\star}$ and SFR:
\begin{equation}
\label{eq:fit}
\begin{array}{rl}
12+{\rm log(O/H)}=&8.90+0.37m-0.14s-0.19m^2\\
            &+0.12ms-0.054s^2\\
\end{array}
\end{equation}
where $m$=log(M$_{\star}$/M$_{\odot}$)-10 and $s$=log(SFR).

In our study we analyze only a subsample of those local SDSS galaxies
that, by construction, lie at redshift $z \sim 0.1 $ on a curve in
this 3D space, represented by the combination of the MSR and MZR.

However, we can test whether the ancestors of
these galaxies  lie or not on the hyper-surface defined by the FMR. 
In fact, the 
\citet{mannucci2010} relation has been defined "fundamental"  implying
that, in principle,  it has to be valid at all redshifts.

In Figs. \ref{FMR_SFR} and \ref{FMR_Z01} we compare in the 3D space defined by $M_*$, SFR and metallicity the properties of 
the model galaxies computed at redshift $z=0.1, 0.4, 0.8, 1.5, 2.2$ with the FMR of \citet{mannucci2010}, 
color coded as a function of SFR and redshift, respectively.

Figs. \ref{FMR_SFR} and \ref{FMR_Z01} show that in this 3D space, our galaxies lie on a surface which is in reasonable 
agreement with the FMR determined by \citet{mannucci2010}.

\subsubsection{Downsizing and the  star formation efficiency} \label{SFE_const}
In this Section  we discuss the robustness of our results in accounting for 
galactic downsizing  with respect 
to the SF efficiency, one key parameter of chemical evolution models.  
As discussed in Section \ref{introduction}, an increase of the SFE as a function of mass is often invoked in chemical evolution models
to explain galactic DS, in particular as traced by the integrated stellar [$\alpha$/Fe] abundance of the stellar
populations of local elliptical galaxies \citep{matteucci1994,pipino2011}. 

As introduced in Section \ref{meth}, we assume the relation of \citet{boselli2014} between SFE and galaxy mass,
where in our case the stellar mass is replaced by the infall mass $M_{inf}$ (as done also in \citealt{spitoni2017}). 
However, this assumption is not key for our results, in particular as far as the capability of our models to account for the
age vs stellar mass relation is concerned.  
To show this, we present the results obtained by adopting a constant SFE=1 Gyr$^{-1}$, i.e. independent on the mass of the galaxy. 
In the right panel of Fig. \ref{SFE1}, we show the SFE as a function of the infalling gas mass used here and adapted from the
\citet{boselli2014} scaling relation compared to a constant value of SFE=1 Gyr$^{-1}$. 
As visible from Fig. \ref{SFE1}, the variable SFE spans a large range, between 0.01 and 4.5 Gyr$^{-1}$.
The relation of \citet{boselli2014} presents SFE$\sim 1$ Gyr$^{-1}$ at mass values $\sim 10^{10.5}~M_{\odot}$, with a factor $\sim 4.5$ 
variation between the value at this mass up to the one of galaxies ten times more massive. 
We also note that low mass systems can have SFE values which are lower even by a few orders of magnitude.

In Fig. \ref{trus_sal_S1}  we present the galactic age-stellar mass relation at different redshifts  
computed for the galaxies that at redshift $z=0.1$ are part of the star-forming main sequence and of the MZR,
but assuming a constant SFE of 1 Gyr$^{-1}$, a value which is typical for chemical evolution models of Milky Way-like galaxies.

A clear DS in the galaxy models is still evident even assuming a constant SFE, in that on average, 
the larger the stellar mass, the older their stellar populations. 
This confirms that our main result, namely that galaxies that lie on both the MZR and MSR 
will also necessarily be characterised by a downsizing in their 
stellar populations, is independent from the adopted star formation efficiency. 
To our knowledge, this is the first time that this result is derived theoretically.

\subsubsection{The effect of a fully T$_{\rm e}$-based abundance scale for galaxies}\label{sec_curti}

 \citet{curti2017} presented new empirical calibrations for strong-line diagnostics of gas phase metallicity in local star forming galaxies by uniformly applying the T$_{\rm e}$  method over the full metallicity range probed by the SDSS. 
Moreover, a MZR for local SDSS galaxies obtained with the T$_{\rm e}$-based method was presented. 

The new fit of the median MZR for the SDSS sample is the following:  
\begin{equation} 
\label{eq_curti}
\text{12 + log(O/H)} = \text{Z}_{0}  - \gamma/\beta * \text{log}\bigg(1 + \bigg(\frac{\text{M}_{\star}}{\text{M}_{0}}\bigg)^{-\beta} \bigg),
\end{equation}
where $\text{Z}_{0}=8.793$,  $\text{log}(\text{M}_{0}/\text{M}_{\odot} )=10.02$, $\gamma=0.28$ and  $\beta=1.2$. 
From the left panel of Fig. \ref{SFE1}, it is clear that the slope and overall normalisation of the MZR are sensitive to
the calibration method (see also \citealt{kewley2008}), and that the relation obtain by \citet{curti2017}
is flatter and generally characterised by lower metallicity values than the one of \citealt{kewley2008}. 

In this Section, we test whether the assumption of a  purely T$_{\rm e}$-based MZR also affects the main results presented above. 

Figures \ref{curti_34},  \ref{curti_lambda} and \ref{curti_trus}
show results qualitatively similar to the ones
of Figs, \ref{z01_12p},  \ref{D_lambda_masses} and \ref{trus_sal}, respectively.
The MZR of \citet{curti2020} produces stronger winds than \citet{kewley2008} MZR (Figs. \ref{curti_34} and \ref{curti_lambda}).
This result can be ascribed to the smaller oxygen abundances 
of the galaxies which follow the \citet{curti2020} MZR, 
as in the framework of the present work, 
it is by means of stronger winds that galaxies can have a lower metal content than the
models computed when the \citet{kewley2008} MZR was assumed. 

Fig.~\ref{curti_trus} shows that also galactic
downsizing is preserved when the \citet{curti2020} MZR is used, with a predicted age-mass
relation very similar to the one of Fig. \ref{trus_sal}.

 \subsection{Forward evolution}

\subsubsection{The MZR by \citet{maiolino2008} and MSR at $z=2.2$}\label{maiolino2_2}
In this work we are also interested in the evolution of galaxies that at high redshift
are part of both the main sequence and of the MZR.
As for the MZR, the observational constraint we discuss here is the one by \citet{maiolino2008}, 
whereas the star-forming MSR is the one by \citet{pearson2018} for galaxies observed in the redshift
range $1.8 \leq z \leq 2.3$.

In the middle panel of Fig. \ref{SFE1} we compare the local MSR
with the one at high redshift by  \citet{pearson2018}.
Large differences characterize the two sequences, in particular as far as the zero-point of the relation is concerned.
An offset by at least one order or magnitude in SFR is visible for instance at stellar masses log(M$_{\star}$/M$_{\odot}$) $\sim$ 11,
whereas any variation in the slope is much less appreciable.
Such a difference in normalization is expected from the evolution of the cosmic star formation rate density 
(e. g., \citealt{madau2014}) and comparable to what found in other studies (e. g., \citealt{santini2017}). 
The MZR considered in this section (solid thick red line in the left panel of Fig. \ref{SFE1})  
is steeper than the local one (solid thick blue line) and, as discuss by \citet{maiolino2008}, this might imply
different evolutionary time-scales in galaxies of different masses.

\begin{figure}
\begin{centering}
\includegraphics[scale=0.35]{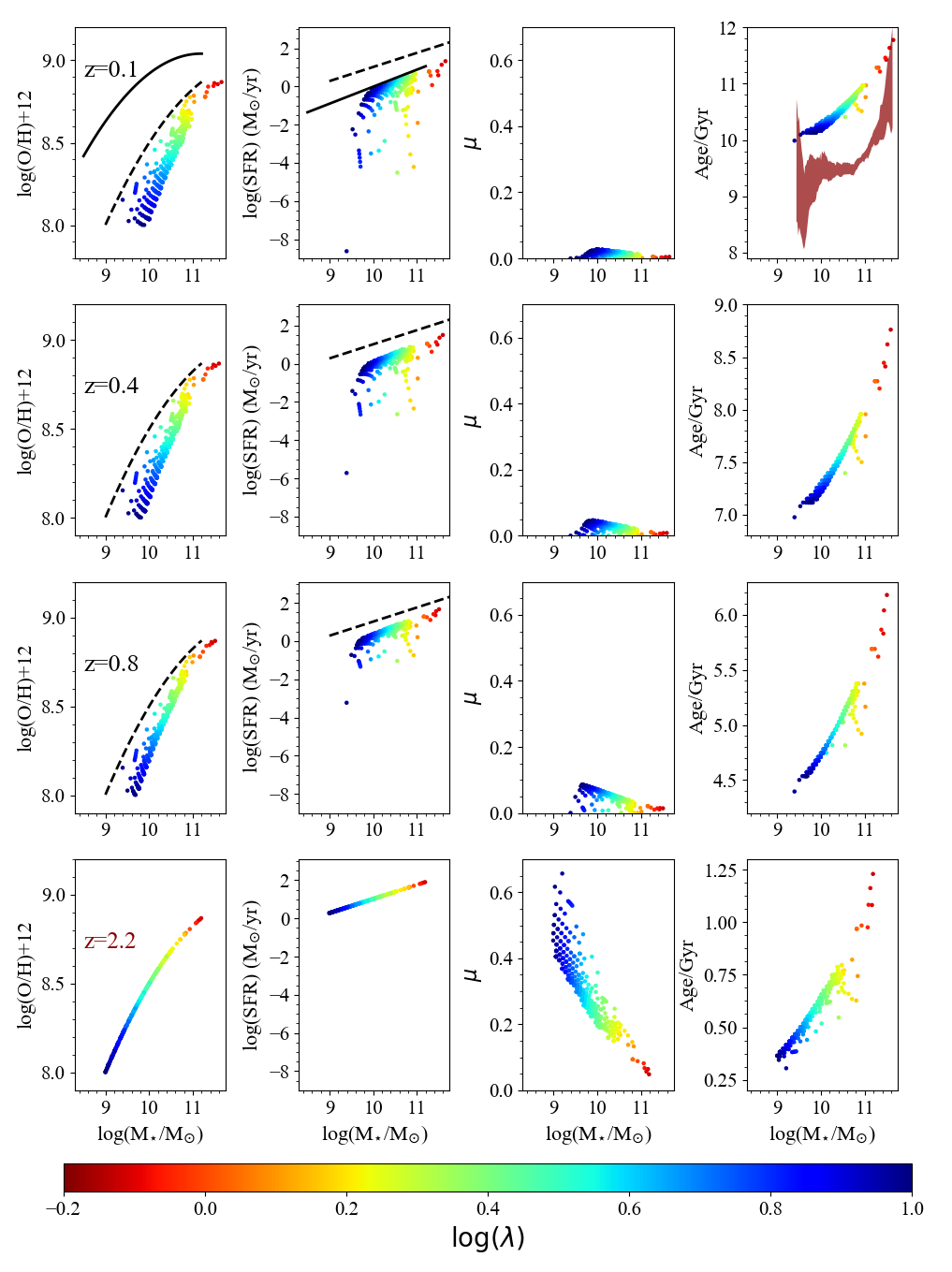}
\caption{Forward evolution from $z=2.2$ to $z=0.1$ of the galaxies which obey the analytical fit of the observed 
MZR at $z\sim 2.2$  by \citet{maiolino2008} and the  main sequence of star-forming galaxies as derived by \citet{pearson2018}. 
Starting from the left, in the first, second and third column the evolution of the MZR, SFR vs. stellar mass and gas fraction vs stellar mass are shown,
respectively. In the last column, the age-stellar mass relation is shown.
The colour coding indicates the loading factor parameter $\lambda$.   The black dashed lines in  the panels of the first column  and  in each  SFR-log(M$_{\star}$/M$_{\odot})$ plot, show the  MZR at redshift $z \sim 2.2$ of \citet{maiolino2008}  and the MSR derived  in high redshift star forming galaxies  by \citet{pearson2018}, respectively.  
 The thick, black solid lines in the first two top-left panels 
  are the MZR at redshift $z \sim 0.1$ of \citet{kewley2008} and the MSR of \citet{peng2010}, respectively. In the upper-right panel with the dark red shaded area we also include the age-stellar mass relation for the local passive galaxies by \citet{trussler2020}.  }
\label{z2_2_pearson}
\end{centering}
\end{figure}

In Fig. \ref{z2_2_pearson} we show the 'forward' (i. e. evolved towards lower redshifts, corresponding to later epochs) 
evolution of the galaxies which at $z=2.2$ obey the MZR and MSR, computed at $z=2.2$, $z=0.8$, $z=0.4$, $z=0.1$.

\begin{figure}
\begin{centering}
\includegraphics[scale=0.4]{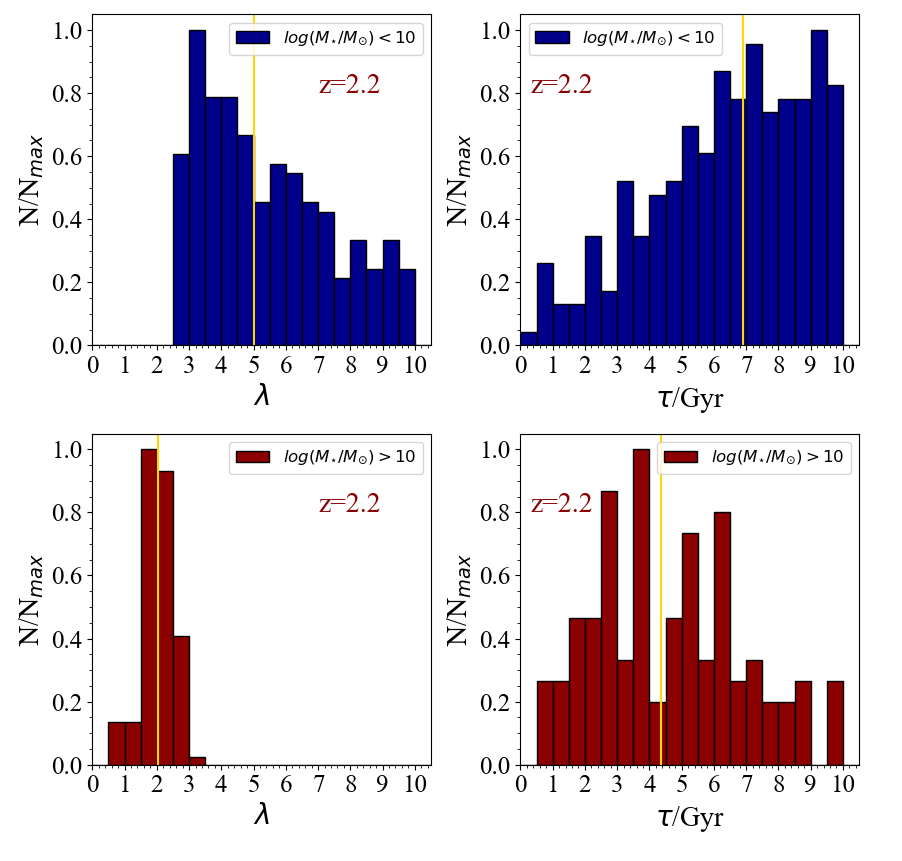}
\caption{Distribution of the predicted galaxies  of Fig. \ref{z2_2_pearson} in terms of the loading factor parameter $\lambda$
  (left panels) and the infall time-scale parameter $\tau$  (right panels), computed at redshift  $z= 2.2$  for different stellar mass bins.}
\label{maiolino22_distrib}
\end{centering}
\end{figure}

 The large SFR values required by the high redshift MSR and the associated  small oxygen abundances of the  MZR relation   force the galactic systems to  be subject to stronger winds  and  assembled on shorter time-scales compared to  the galaxies  presented in Section \ref{backward_sec}  (see Fig. \ref{maiolino22_distrib}).
Hence,  galaxies suffer strong depletion  of gas and   evolve passively at  the redshifts $z<2.2$.  In Fig. \ref{z2_2_pearson}, it is evident that  the gas fraction drops dramatically   from redshift $z=2.2$ to redshift $z=0.8$.  The most massive objects show high stellar masses (built up in a short time) and high metallicity and must have consumed their reservoir
of gas soon after redshift $z=2.2$. After that, the buildup of their stellar mass and  metals was already complete,
and they have undergone little evolution at later times. Less massive galaxies, characterized by stronger winds (see Fig. \ref{maiolino22_distrib}), present a more sensible evolution, which manifests in
a steepening of the MZR towards recent times. 
In conclusion,  the forward evolution of the single galaxies  in the MZR relation is characterized by an "horizontal" evolution at roughly constant metallicity.

\begin{figure*}
\begin{centering}
\includegraphics[scale=0.37]{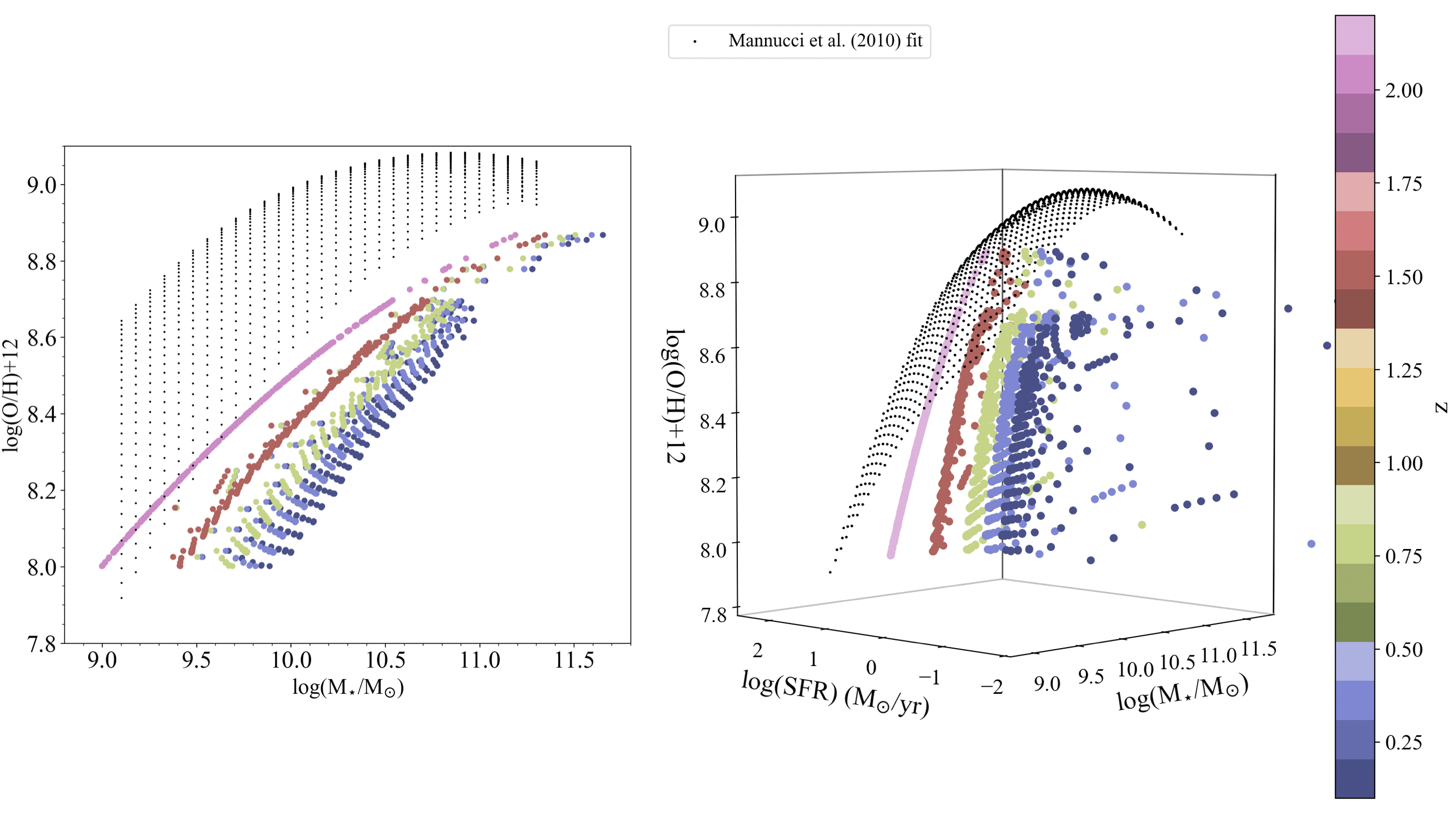}
\caption{ As in Fig. \ref{FMR_Z01} but for galaxies which at $z= 2.2$ follow the MZR of \citet{maiolino2008} and
    the MSR derived by \citet{pearson2018}. The big solid circles are computed  at redshifts $z= 2.2$, $z= 1.5$, $z= 0.8$,  $z= 0.4$, and  $z= 0.1$ and
    are colour-coded as a function of the redshift value. Black dots are as in Fig. \ref{FMR_Z01}. } 
\label{FMZ_z22}
\end{centering}
\end{figure*}

 The passive evolution of these galaxies can be also inferred  by the  their age-stellar mass relation  at redshift $z=0.1$, which is   similar to the one followed by local passive galaxies of \citet{trussler2020}.
However,  passive galaxies in \citet{trussler2020} have much larger stellar metallicities  than simulated  galaxies at redshift $z=0.1$. 
In principle, these galaxies could have been part   of the local passive sequence of \citet{trussler2020} as due to a quenching mechanism, such as the one  proposed by \citet{peng2015}: e.g.  by "strangulation" (see their lower panel in Fig.1). After the beginning of the "strangulation"  no outflow and no gas accretion affect the galactic system and the   star formation can continue with the gas available in the galaxy until it is completely used up. During this phase the gas metallicity increases substantially because of the lack of dilution from inflowing gas and gas loss.

As already underlined by \citet{maiolino2008}, galaxies belonging to the MZR observed at different redshifts should not be seen as an evolutionary sequences. 
In fact, Fig. \ref{maiolino22_distrib} clearly shows that at redshift $z=2.2$ these galaxies are characterized by stronger winds compared to
the galaxies studied in Section \ref{MZ_01}. This is a result of imposing that they also belong to the MSR which, at $z=2.2$, on the entire stellar   mass range 
is characterised by larger SFR values than the one low redshift. 
In order to present at the same time low metallicities and high SFR values, it is clear that 
galaxies at $z=2.2$ must have suffered much stronger stellar winds than galaxies belonging to the MZR and MSR at lower redshifts. 

 In Fig. \ref{FMZ_z22} we compare the properties of the galaxies 
which at redshift $z \sim 2.2$ follow the MZR and MSR relations with the FMR of \citet{mannucci2010}.
The models shown in  Fig. \ref{FMZ_z22} are computed at redshifts $z=0.1, 0.4, 0.8, 2.2$, and color-coded as a function of $z$. 
The loci occupied by the model galaxies partially overlap with the FMR only at $z=2.2$, whereas at lower redshifts 
the fall off the surface traced by the black points, which represent the analytical fit to the FMR presented in Sect. \ref{sec_FMR}.
This due to the fact that galaxies which at $z=2.2$ follow both MZR and MSR become passive at later times, as shown by the progressive decline 
of their SFR values.

\begin{figure}
\begin{centering}
\includegraphics[scale=0.35]{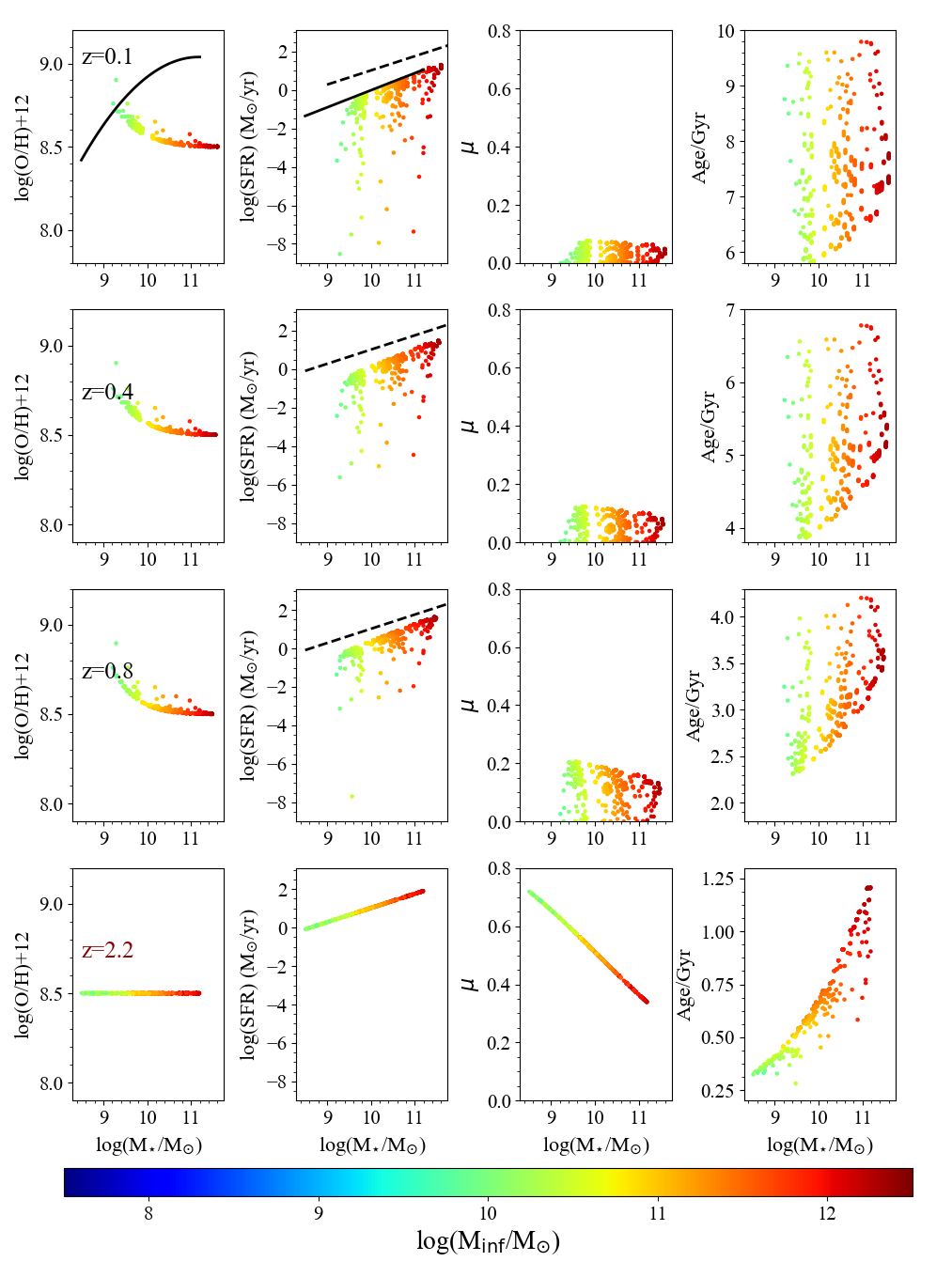}
\caption{ 
Forward evolution from $z=2.2$ to $z=0.1$ of the galaxies which obey the analytical fit of the 
MZR at $z\sim 2.2$  observed by \citet{mignoli2019} and the main sequence of star-forming galaxies as derived by \citet{pearson2018}. 
Starting from the left, in the first, second, third and fourth column the evolution of the MZR, SFR vs. stellar mass, gas fraction vs stellar mass
and age-stellar mass relation are shown, computed at redshifts 
$z= 2.2$, $z= 0.8$, $z= 0.4$, and  $z= 0.1$ (from bottom to top)
The color coding stands for the infall gas mass value.
 The thick, black solid lines in the first two top-left panels 
  are the MZR at redshift $z \sim 0.1$ of \citet{kewley2008} and the MSR of \citet{peng2010}, respectively.
The black dashed lines  in each  SFR-log(M$_{\star}$/M$_{\odot})$ plot stand for the MSR derived  in high redshift star forming galaxies  by \citet{pearson2018}.   }
\label{z22_16_mignoli}
\end{centering}
\end{figure}

\subsubsection{The MZR by \citet{mignoli2019} and  MSR at $z=2.2$}\label{mignoli2_2}
In the narrow line regions (NLR) of a sample of 88 \civ-selected objects containing type 2 AGNs, 
\citet{mignoli2019} measured at $z\sim 2.2$ a nearly flat  MZR, set at the value of 12+log(O/H) $\sim$ 8.5 dex. 
It is worth mentioning that the computed metallicity of the NLR in the selected type 2 AGNs is based on calibrations from photoionization models of various rest-UV diagnostic line ratios. These methods are quite different from those adopted in "normal" star-forming galaxies as in \citet{maiolino2008}, which involve rest-optical lines and a combination of T$_{\rm e}$-measurements and photoionisation modelling, hence the comparison between the two should be taken with caution.

As suggested by \citet{mignoli2019}, one of the possible reasons for a flat MZ relation 
is a selection effect in the sample. 
The high-ionization \civ\   emission line can cause culling of massive and dust-free hosts,
not representative of the star-forming galaxy population at these redshifts.
Another possibility is that the metallicity  of the NLR is not a good proxy of that of the host galaxy,
because an AGN can strongly affect not only the ionization properties of the gas, but also the circulation 
of metals in the ISM, e. g. by means of a strong outflow. 

\begin{figure}
\begin{centering}
\includegraphics[scale=0.4]{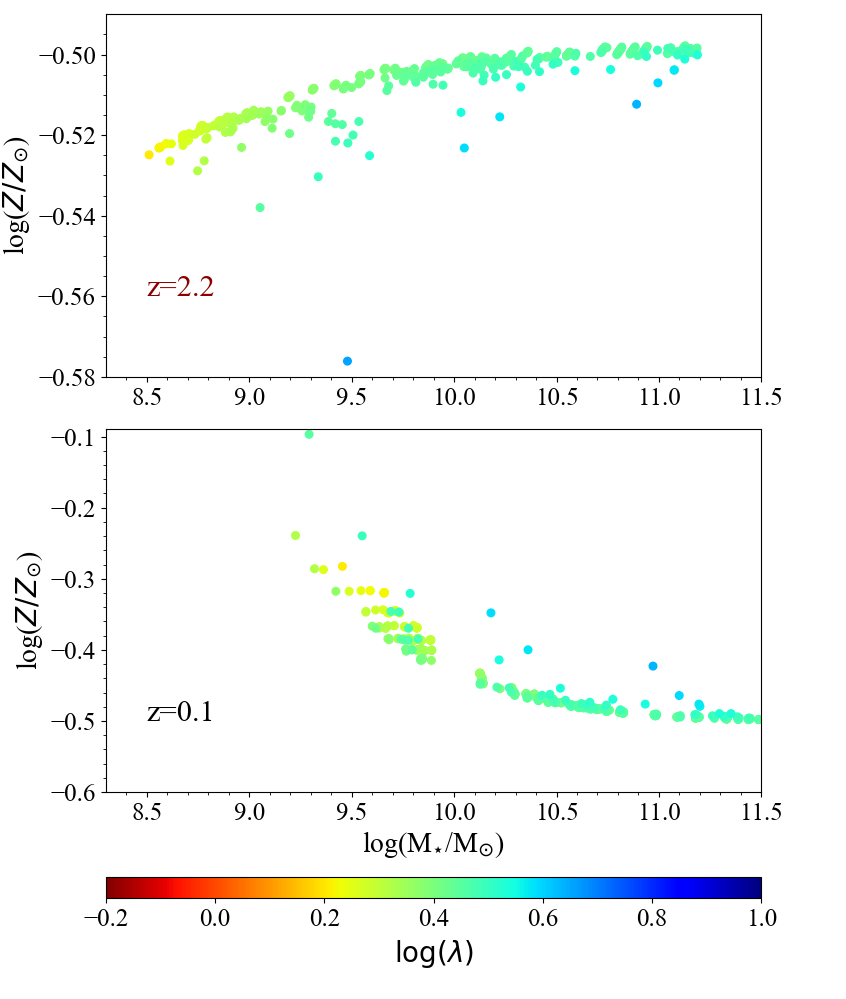}
\caption{Stellar mass-metallicity relation for the galaxies which follow the \citet{mignoli2019} MZR and the high redshift star forming main sequence by \citet{pearson2018} computed at  redshift $z=2.2$ (upper panel) and $z=0.1$ (lower panel). }
\label{zstar_mignoli}
\end{centering}
\end{figure}

We make the hypothesis that the galaxy sample of \citet{mignoli2019} include average star-forming galaxies. 
By means of our models, we have computed  the forward evolution of the galaxies that at $z$=2.2 belong to the star-forming
MSR of \citet{pearson2018} and also present a flat MZR as observed by \citet{mignoli2019}, as shown in Fig. \ref{z22_16_mignoli}. 

We have used the model with a constant SFE, fixed at the value of 1 Gyr$^{-1}$ as discussed earlier. 

From Fig. \ref{z22_16_mignoli}, we note that the required infall masses are larger than the
galaxies that belong to the local MZR and the MSR (see Fig. \ref{z01_12p_minf}).

A high oxygen abundance has to  be already in place at high redshift even for galaxies with low stellar masses, and 
large infall masses are naturally accompanied by large metal production.\\
As for high stellar mass systems, the MSR imposes large SFR values ( $1<\log({\rm SFR}/M_{\sun}, {\rm yr}^{-1})<2$).
In order to maintain a high star formation level, large reservoirs of gas (i.e. large infall masses) are required.

 In Fig. \ref{z22_16_mignoli},  we  note that the  mass-weighted ages versus stellar mass  relation (computed ad different redshifts) shows a larger  spread compared the one presented in Fig. \ref{z2_2_pearson} where at high redshift has been imposed the MZR of \citet{maiolino2008}. We tested that this difference is mainly  due to the fact that in the last  case   galaxies with the  same infalling mass present a similar SFH, and consequently they have the same  mass-weighted age. In contrast, the high redshift  flat MZR leads to a wide range of SFH  at a fixed value for the infall mass in the successive evolution phases.

In their subsequent evolution, the galaxies settle onto  an 'inverted'  MZR, as already visible at $z=1$. 
This occurs because, owing to their large SFR values, the  system with the largest stellar masses become soon passive, hence 
their metal content does not change significantly at later times. 

Low-mass galaxies are characterised by lower SFR values and by an already high metallicity.
Their low SFR values and their relatively large infall masses allow them to maintain their star formation
activity and to keep producing new metals, with an unavoidable consequent increase of their metallicity. 
In this case, their winds are not sufficient to lead to a significant decrease of their metallicity and to
have them settled on a 'normal' MZR at lower redshifts.

These results might indicate that either our hypothesis regarding that these galaxies are on the MSR is inadequate, or
that one fundamental ingredient is missing in our models which causes a substantial redistribution of metals,
such as the feedback of an AGN. 
At present, investigations are on going in order to constrain the SFR of these systems and to assess which region they
occupy in the SFR-$M_*$ diagram, to be compared  to the MSR of 'normal' star forming galaxies observed at $z \sim 2$.

In Fig. \ref{zstar_mignoli} we explore the evolution of the stellar metallicity for the galaxies of the \citet{mignoli2019} flat MZR. 
This quantity can be computed from our models in  the following way:
\begin{equation}
\left< Z_{\star}(t) \right> =\frac{\int_0^t dt' \, Z(t') \, \psi(t')}{\int_0^t dt'\,\psi(t')}. 
\label{mass-weighted_Z}
\end{equation}
Eq.~\ref{mass-weighted_Z} is by definition the mass-weighted stellar metallicity,
i. e. the mass-weighted average of the metallicity of the stellar populations in each single galaxy (see \citealt{pagel1997}). 
It is interesting to note that at redshift $z=2.2$, the shape of the stellar MZR is much different from the flat relation of the gas-phase metallicity.
The stellar MZR presents more of a standard form, with an increasing behaviour at the low-mass edge and a flattening at
$log (M_{*}/M_{\odot}) \sim 10$. 
This is a consequence of having imposed the MSR, and it is due to the fact that more massive objects
have suffered on average stronger SF episodes at their earliest stages.

However, this relation is much flatter than the relations discussed previously in Sect.~\ref{maiolino2_2}, as 
the variation between the metallicity of the lowest mass galaxies with $log (M_{*}/M_{\odot}) \sim 8.5$ and the plateau is 
of only 0.02 dex.
At redshift $z=0.1$, also the stellar metallicity presents an inverted shape, the with highest metallicity values at the lowest masses,
as already seen in Fig. \ref{z22_16_mignoli}.\\
In summary, in our view the galaxies of a sample like the one of Fig. \ref{z22_16_mignoli} and which build up a flat MZR
do not represent an average sample of  star-forming galaxies at high-redshift, which are expected to follow a steep MZR, as other works have
shown \citep{erb2006,maiolino2008}.
The sample of Fig. \ref{z22_16_mignoli}, in which all galaxies include a Type 2 AGN,
is likely to be composed of systems at various evolutionary stages, and in which the AGN played an important role
in regulating the star formation activity.

 As a final note, another work in which the MZR in galaxies hosting type-2 AGN is 
the one of \citet{matsuoka18}. 
Their sample includes high-z radio galaxies and X-ray selected radio-quiet AGNs at $1.2<z<4.0$ 
with stellar mass values $>10^{10}~M_{\odot}$.
In this mass interval, \citet{matsuoka18} derive a MZR consistent with the one of \citet{maiolino2008}.
However, due to the selection of their sample, their study does not probe the low-mass end of the MZR.
In the future, metallicity measures of more extended datasets of AGN hosts will be
needed to shed more light on the evolution of their MZR, and to further highlight possible differences with the scaling 
relations of star-forming galaxies.

\section{Conclusions}\label{conc}

In this paper we have studied the evolution of the systems which follow 
two fundamental observed scaling relations for star-forming  galaxies, namely the MZR and the MSR. 
Our study was performed by means of  the  analytical, 'leaky-box' chemical evolution model  of
\citet{spitoni2017}. 
In such model, galaxies are assumed to be formed by means of 
accretion of primordial gas, with an accretion rate which follows an exponential
law, and  in presence of  galactic winds. 
In the model, each galaxy is characterized in terms of three model free parameters: the total
infalling mass $M_{inf}$, the loading factor wind parameter
$\lambda$ and the the time-scale of gas accretion $\tau$.

We focused on a given redshift, and we imposed that the galaxies must
follow the observed MZR and the star-forming MSR measured at that particular 
redshift.  First,  we  studied  the  properties  of  the  local
galaxies  by  imposing the MZR and the MSR observed at redshift $z=0.1$, and
analyzing their ‘backward’ evolution towards higher redshifts. On the
other hand, by imposing the same scaling relations at higher redshift,
we  have computed a ‘forward’ evolution of the galaxies at lower
redshifts and towards the present time.

Our main conclusions can be summarized as follows:

\begin{itemize}

\item By imposing the MZR and the MSR at redshift $z=0.1$, we showed
that the  galaxies already present at redshift $z=2.2$ are
characterized by weak winds and long time scales of gas accretion.
In fact, at low redshift these objects have to maintain a large
reservoir of gas in order to be part of the local MSR and a
high-metallicity, as they now build the high-metallicity plateau of
the MZR.  Moreover, the median of the infall time-scale distribution
in  two  different  bins  of  stellar  masses ($M_{\star}<10^{10} M_{\odot}$
and $M_{\star}>10^{10} M_{\odot}$)  decreases with
decreasing redshift.  This reflects that at each redshift, only the
youngest galaxies can be assembled on the shortest timescales and
still belong to the star-forming MSR.  In the lowest mass bin, a
decrease of the median $\tau$ is accompanied by an increase of the
median $\lambda$ value.  This implies that systems which have formed
at more recent times will need to eject a larger amount of mass to
keep their metallicity at low values.

\item 
 Imposing both the MZR and the MSR naturally leads to an increasing relation between the mass-weighted age and the stellar mass which is qualitatively in agreement with the relation observed in local SF galaxies obtained with independent determinations. We note though that a quantitative agreement is difficult to assess because different observational estimates of the age-mass relation are sensitive to the assumption on SFH, SSP models and technique used.
 To our knowledge, such a direct connection between
the MZR, MSR and downsizing was never shown before in any theoretical
study. Moreover, no successful attempt has been performed
so far to simultaneously reproduce these fundamental scaling
relations within a cosmological framework, mostly because of an early overproduction of
low-mass galaxies, which in such models nowadays are dominated by too old stellar populations 
(\citealt{calura2014}, \citealt{somerville15}).

\item 
In a  3D space formed by SFR, oxygen abundances and stellar mass, 
the ancestors of the galaxies that follow  the observed MZR and the star-forming MSR at $z=0.1$ (as studied at $z\le2.2$) 
lie on a hyper-surface which is in good agreement with the fundamental metallicity relation of \citet{mannucci2010}. 

\item We studied  the evolution of galaxies that at  high redshift are
part of both the observed MSR and of the MZR.  By imposing at redshift
$z=2.2$ the MZR of \citet{maiolino2008}, we found that the large
SFR values required by the high redshift  MSR  forces  most  of  the
galaxies  to evolve  passively at  redshifts  smaller  than
$z=2.2$. In fact,  the  most  massive  objects show high stellar
masses (built up in a short time) and high metallicity and must have
consumed their reservoir of gas soon after leaving  the MZR. Less
massive  galaxies,  characterized  by  stronger  winds, present a
slower evolution, which manifests in a progressive steepening of the MZR
towards recent times. 
In order to present at the same time low metallicities and high SFR values, it is clear that 
galaxies at $z=2.2$ must have suffered much stronger stellar winds than galaxies belonging to the MZR and MSR at lower redshifts. 
Therefore, our findings confirmed that the observed MZRs and MSRs at different redshifts should not be interpreted as evolutionary sequences (see also \citealt{maiolino2008} and  \citealt{lian2018ev}).
     
\item In our analysis, we have calculated also the 'forward' evolution of high
redshift objects which at $z\sim 2$ are characterized by a flat MZR, as observed  by
\citet{mignoli2019}. The sample by \citet{mignoli2019} includes galaxies with
a Type 2 AGN and selected from their \civ\ emission and for which the star formation rate is unknown.  
In this particular case,  even low-mass galaxies have to present a high metallicity, already  at early times.
Also in this case, the system with the largest stellar masses become soon passive, whereas lower mass galaxies
can still keep their star formation activity, with a further increase of their metallicity and despite
their winds which, even if stronger than more massive systems, are not sufficient to deplete significantly their metal reservoirs. 
The implication is that, in their subsequent evolution, the galaxies settle onto  an 'inverted'  MZR. 
These results might indicate that one fundamental ingredient is missing in our model, 
which causes a substantial redistribution of metals within each single galaxy, such as the feedback of an AGN.
Another possibility is that at present, these systems do not follow the MSR. 
At present, investigations are on going in order to constrain the SFR of these
systems and to assess which region they occupy in the SFR-$M_{\star}$ diagram.

\end{itemize}

\section*{Acknowledgement}
The authors thank the referee for the useful comments, which have certainly improved the paper. We thank G. Zamorani for useful suggestions. 
Funding for the Stellar Astrophysics Centre is provided by The Danish National Research Foundation (Grant agreement no.: DNRF106).
E. Spitoni also acknowledges funding from the INAF-OAS visitor program. 
E. Spitoni and V. Silva Aguirre acknowledge support from the Independent Research Fund Denmark (Research grant 7027-00096B).
V. Silva Aguirre acknowledges support from VILLUM FONDEN (Research Grant 10118).  
FC acknowledges support from grant PRIN MIUR 2017 -
20173ML3WW\_001 and from the the INAF
Main-Stream (1.05.01.86.31).

\bibliographystyle{aa} 
\bibliography{disk}
\appendix
\section{Model results with the \citet{chabrier2003} IMF} \label{section_chab}

In this Appendix we present our model results similar to what discussed in Section \ref{backward_sec}
but computed assuming a \citet{chabrier2003} IMF. 
We still consider as constraints 
the observed  MZR and star-forming main sequence  at redshift $z=0.1$, and we present the 'backward' evolution
of some of the properties that build these two relations, including the gas fraction and age-mass relation. 

As mentioned in Section \ref{ES17} , the \citet{chabrier2003} IMF is richer in massive stars than the Salpeter IMF.
If we assume the compilation of  stellar yields presented by \citet{romano2010}, 
the oxygen yield per stellar generation computed with a \citet{chabrier2003} IMF is
by a factor of $\sim$ 2.3 larger than the value computed with the \citet{salpeter1955} IMF. 

\begin{figure}
\begin{centering}
\includegraphics[scale=0.31]{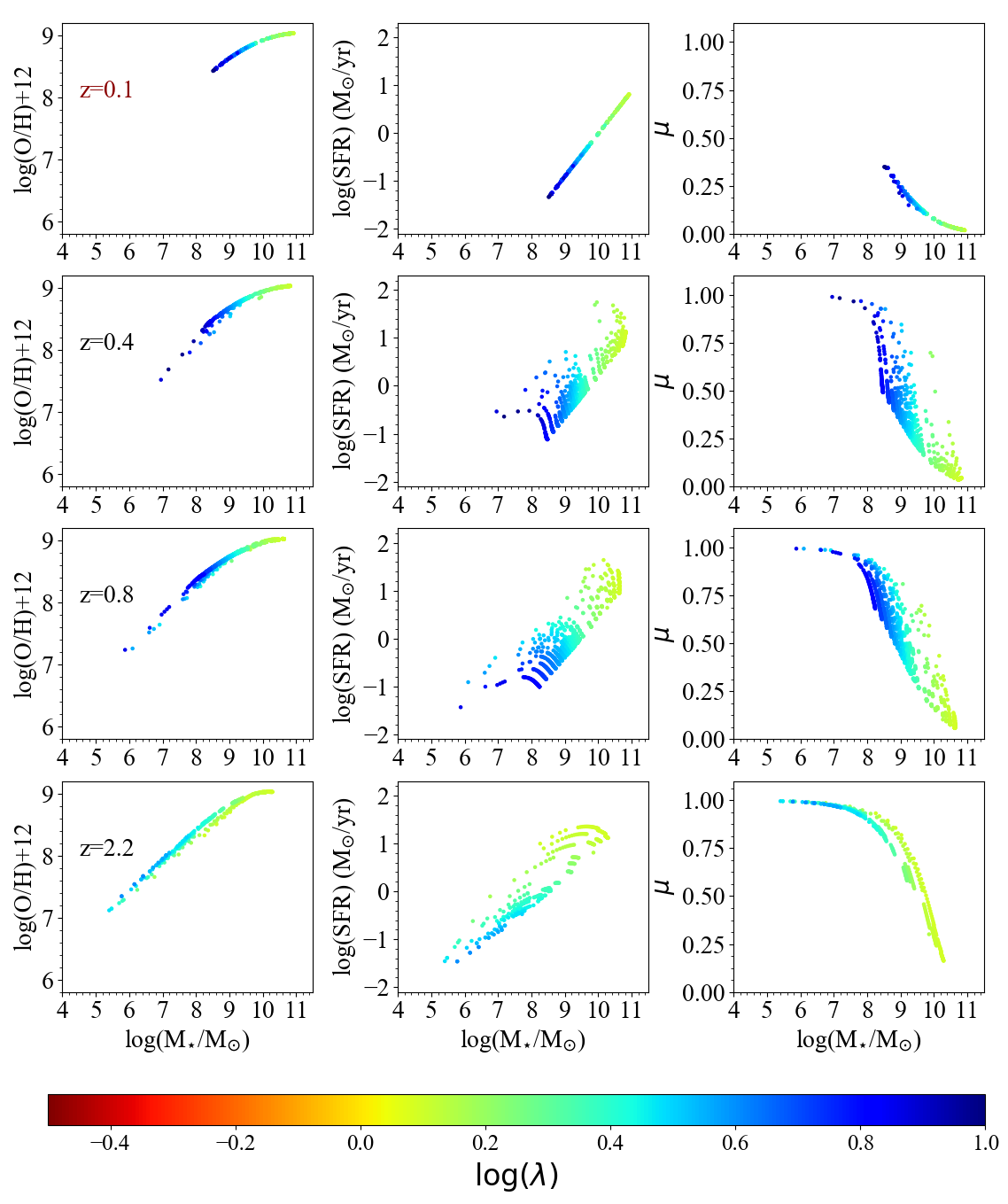}
\caption{
 Backward evolution from $z=0.1$ to $z=2.2$ of the galaxies which obey the analytical fit of the observed
 MZR at $z\sim 0.1$ and the local main sequence of star-forming galaxies as in Fig. \ref{z01_12p},
 but computed assuming a \citet{chabrier2003} IMF.
In the first, second and third column we show the evolution of MZR, SFR vs stellar mass and gas fraction vs stellar mass, respectively.  
The colour coding indicates the loading factor parameter $\lambda$.}
\label{4_3_chab}
\end{centering}
\end{figure}

In Fig. \ref{4_3_chab} we show the backward evolution of the MZR (first column), SFR-mass relation
and gas fraction vs stellar mass relation for galaxies which at redshift  $z=0.1$ lie on the MZR and on the main sequence.

\begin{figure}
\begin{centering}
\includegraphics[scale=0.32]{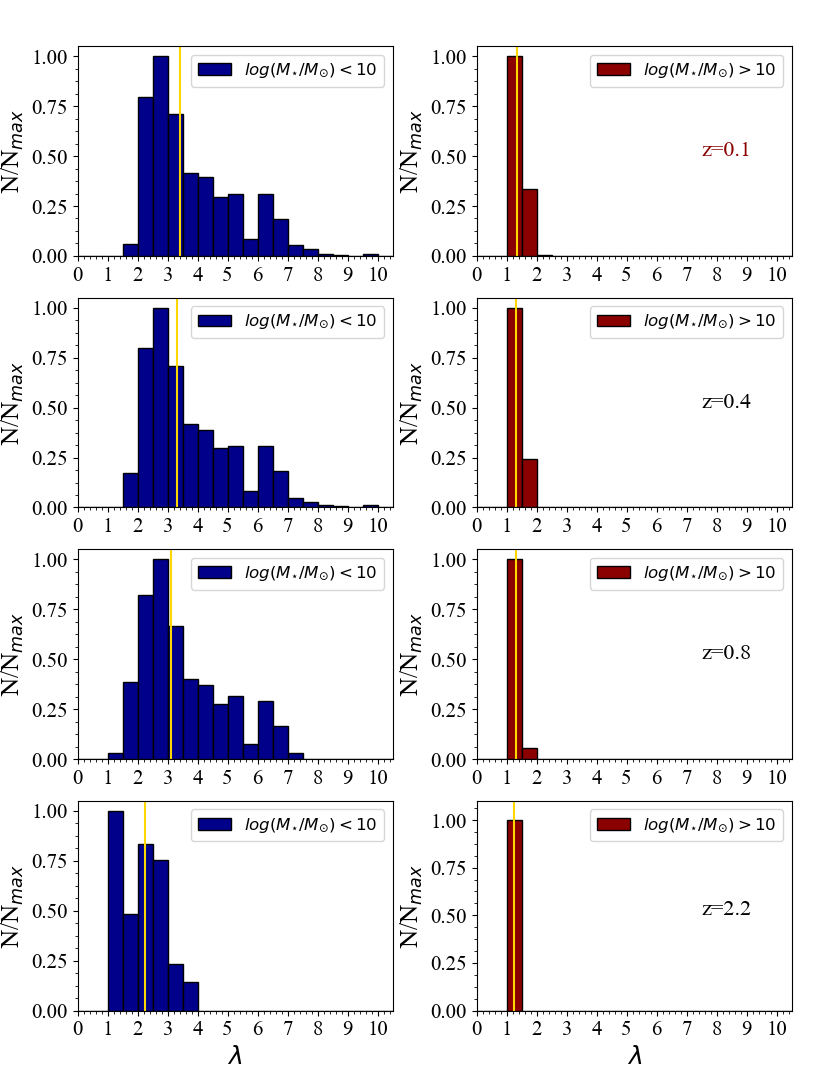}
\caption{Distribution of the loading factor parameter $\lambda$ for the galaxies of Fig. \ref{4_3_chab} 
  computed at redshifts $z=2.2$, $z=0.8$, $z=0.4$, and $z= 0.1$ for 2 different bins of stellar mass (left: $log(M_{\star}/M_{\odot})< 10$;
  right: $log(M_{\star}/M_{\odot})> 10$). The yellow vertical lines indicate the median values of each distribution.
  }
\label{D_l_chab}
\end{centering}
\end{figure}

Compared to \citet{salpeter1955} case, in Fig. \ref{4_3_chab} we note that  the sequence of the first galaxies which have appeared at redshift $z=0.8$ 
is more populated and more extended towards low stellar mass values. 
This is an obvious consequence of the higher yield for stellar generation, which leads to a more efficient chemical enrichment in all galaxies,
and hence at any time galaxies with a smaller stellar mass can show the same oxygen abundance.
Moreover, in Fig. \ref{D_l_chab} we see that the \citet{chabrier2003} IMF requires stronger winds compared to the \citet{salpeter1955} as reported in Fig. \ref{D_lambda_masses}, 
confirming previous results by \citet{spitoni2017}. 

\begin{figure}
\begin{centering}
\includegraphics[scale=0.32]{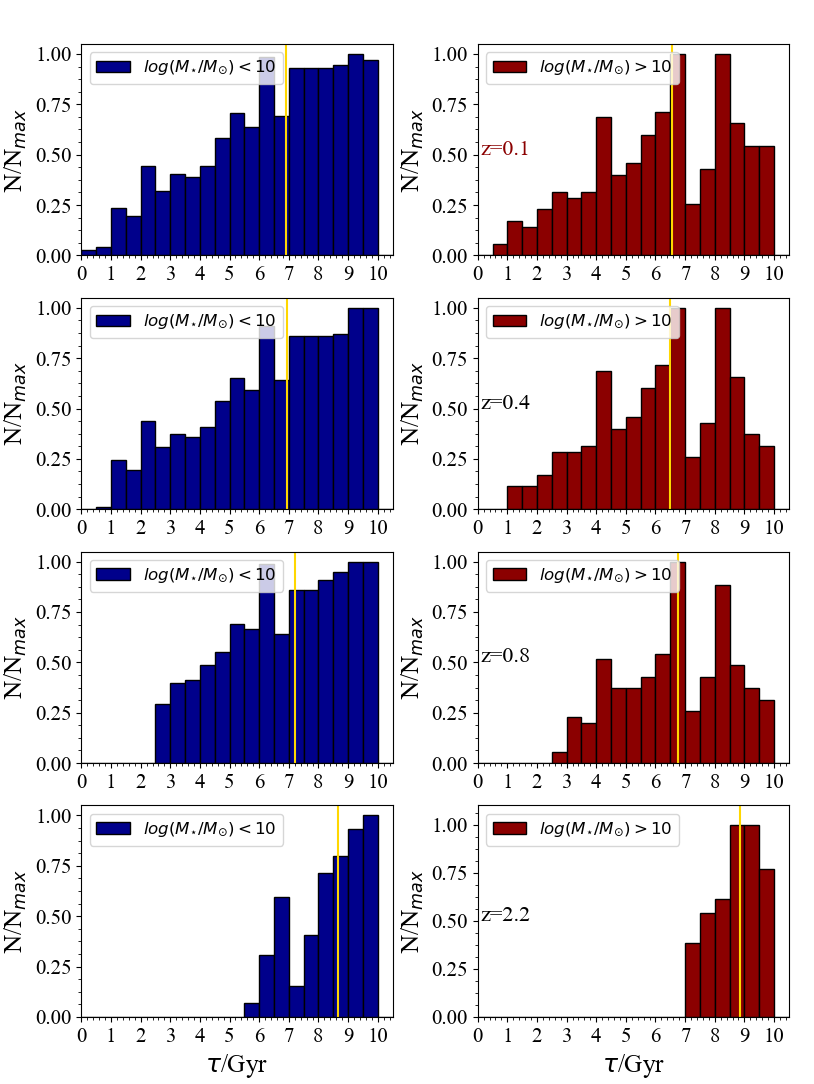}
\caption{As in Fig. \ref{D_l_chab}, but in each panel the plotted distribution is for the infall time-scale parameter $\tau$.}
\label{D_t_chab}
\end{centering}
\end{figure}

\begin{figure}
\begin{centering}
\includegraphics[scale=0.45]{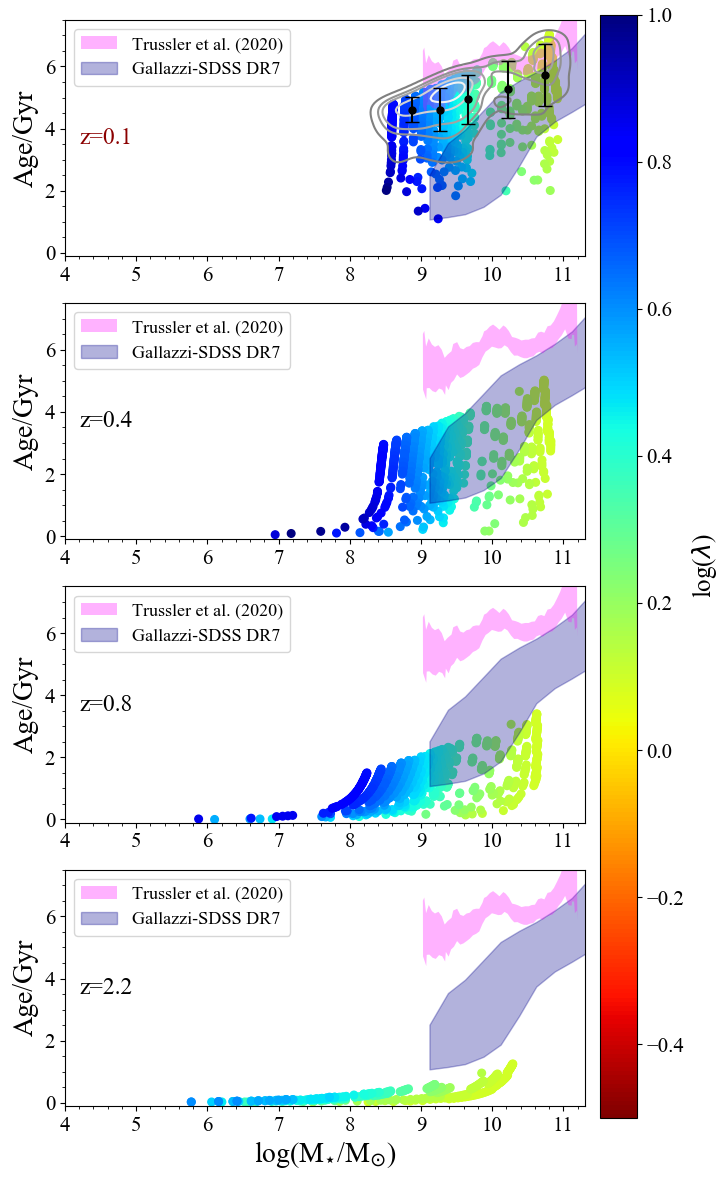}
\caption{Backward evolution of the mass-weighted age versus stellar mass relation for the galaxies of Fig. \ref{4_3_chab}, 
computed adopting a \citet{chabrier2003} IMF.}
\label{trus_chab}
\end{centering}
\end{figure}

\begin{figure}
\begin{centering}
\includegraphics[scale=0.32]{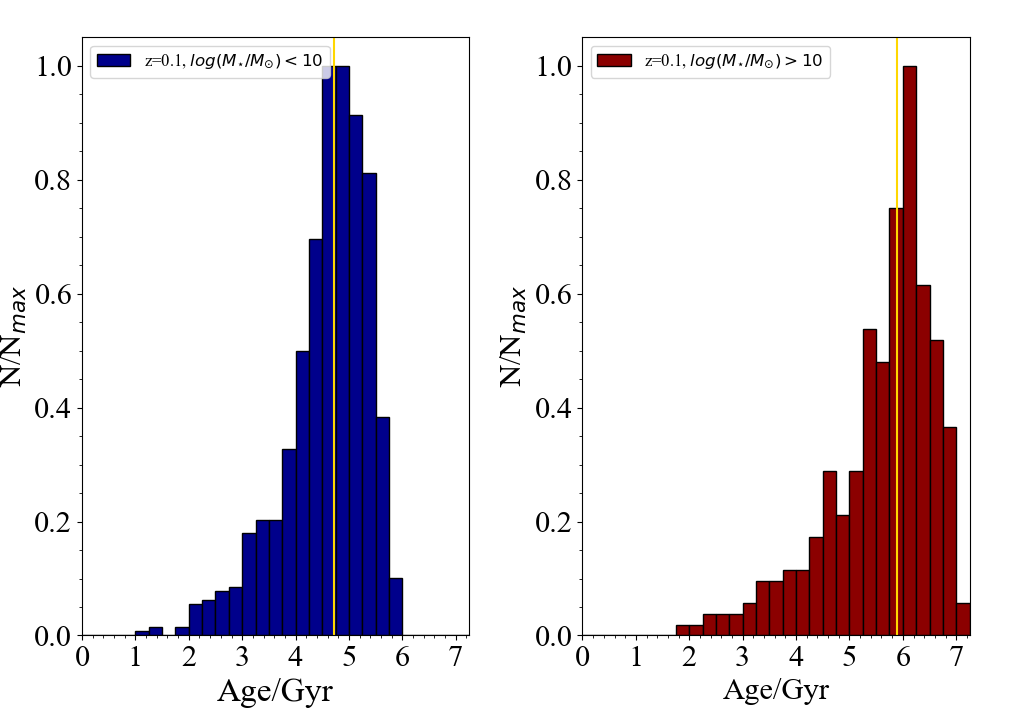}
\caption{Age distribution for star-forming galaxies which are on the MZR at $z=0.1$ in two different mass bins as in Fig.~\ref{age_D},
  computed assuming a \citet{chabrier2003} IMF.}
\label{chab_age}
\end{centering}
\end{figure}

The evolution of the age-mass relation for our galaxies 
is shown in Fig. \ref{trus_chab} and the predicted  age distribution computed at redshift $z=0.1$ in two different redshift bins is shown in Fig. \ref{chab_age}. 
Both figures show that on average, the \citet{chabrier2003} IMF produces older galaxies than the \citet{salpeter1955} IMF. 
 
This result seems to be in contrast with the ones obtained of \citet{spitoni2017} in which, by using as constraint for star-forming galaxies only the local MZR, 
the \citet{chabrier2003} IMF produced younger galaxies than the \citet{salpeter1955} IMF.
This tension is only due to the fact that in this case we use also the MSR as further observational constraint for local star-forming  systems. 

\begin{figure}
\begin{centering}
\includegraphics[scale=0.34]{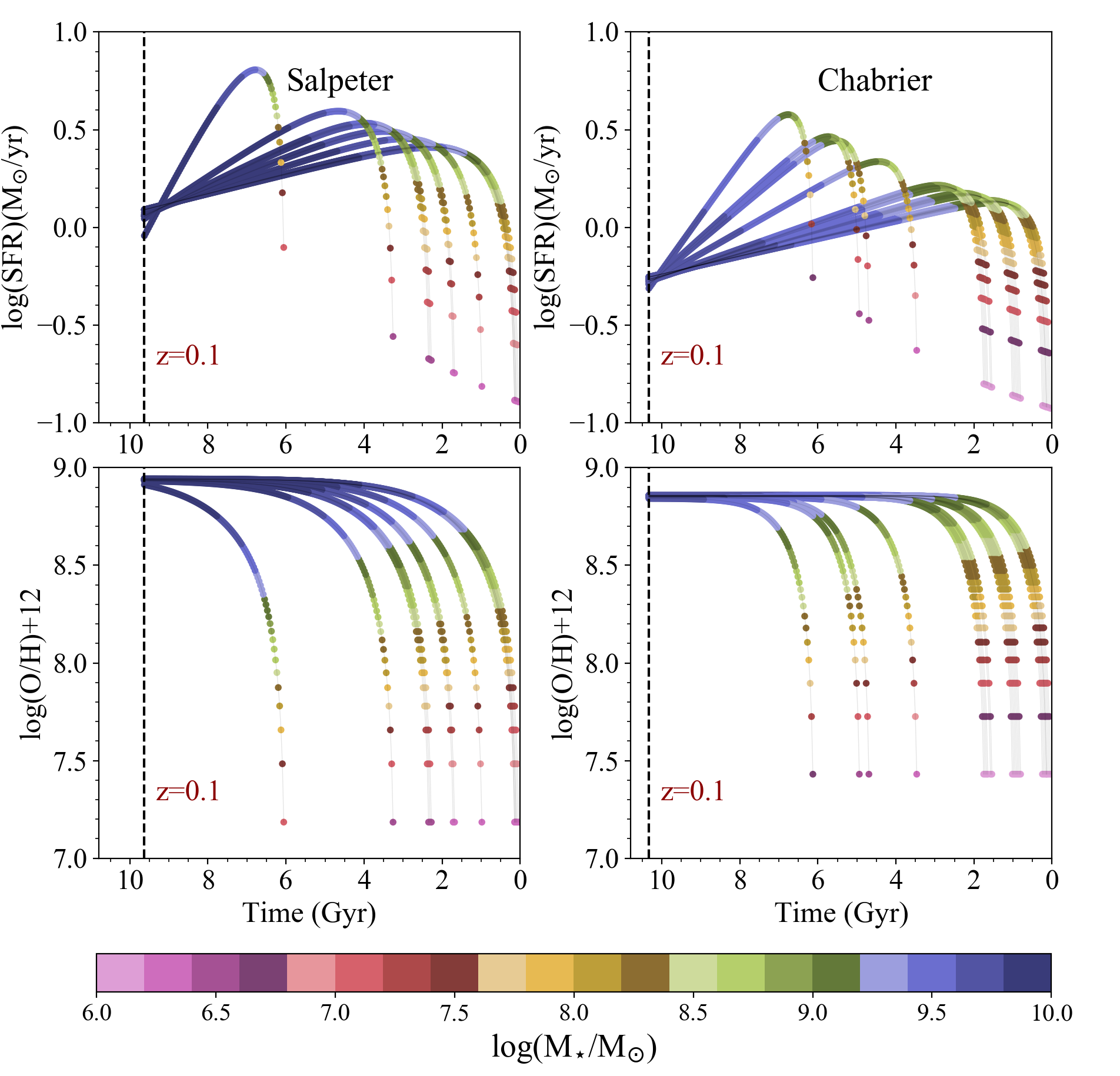}
\caption{Evolution of star formation rate (upper panels) and oxygen abundance (lower panels)  as a function of time  for the galaxies
with infall mass $M_{inf}= 10^{10.5} M_{\odot})$ which lie on the MZR of \citet{kewley2008} and on the MSR as derived by \citet{peng2010},
computed assuming a \citet{salpeter1955} IMF (left panels) and a \citet{chabrier2003} IMF (right panels).  The colour coding stands for the stellar mass.}
\label{10_5_time}
\end{centering}
\end{figure}

In Fig. \ref{10_5_time} we compare 
the temporal evolution of the oxygen gas-phase abundances and of the SFR
computed with two different IMFs.
These quantities have been computed for a set of model galaxies which locally lie on both the  MZR and on the MSR
and which are 
characterised by a constant infall mass, fixed at the value of $M_{inf}=10^{10.5}$ M$_{\odot}$ 

As expected, the metallicity of the galaxies with a \citet{chabrier2003} IMF
reach their saturation value at earlier times than with the  \citet{salpeter1955} IMF.
Stronger winds are required to avoid an excessive increase of their metallicity, with 
the consequence of a significant removal of the gas available for star formation. This implies that, given the same infall mass,
galaxies with a  \citet{chabrier2003} IMF will be forced to evolve at a lower star formation regime.

With  \citet{chabrier2003} IMF  metallicity grows faster but stellar mass grows at a slower pace than with a \citet{salpeter1955} IMF.
In the models shown in Fig. \ref{10_5_time} the SFH is continuous and, because of a more efficient gas removal
in the case of the  \citet{chabrier2003} IMF, 
the only way to have at the present day comparable stellar masses 
will be by means of a more extended SFH.\\
It is worth noting that in the case of Fig. \ref{10_5_time}, the models with a  \citet{chabrier2003} IMF will have
smaller stellar masses than the ones with a Salpeter IMF.
However, the same considerations will hold also for galaxies with the same present-day stellar masses
which, in the case of a  \citet{chabrier2003} IMF, will need to have started forming stars at earlier epochs, and which will
thus present on average older mass-weighted ages.
 With reference to the galaxies  drawn in Fig. \ref{10_5_time} using the \citet{salpeter1955} IMF, the average values   for the stellar mass, the temporal evolution $t_n$ computed at redshift $z=0.1$  and  the associated  mass-weighted age Age$(t_n)$ are respectively: $\log(M_{\star}/M_{\odot})=10.09$, $t_n=8.15$ Gyr and Age$(t_n)= 4.45$ Gyr. On the other hand, with the \citet{chabrier2003} IMF, we have that $\log(M_{\star}/M_{\odot})=9.71$, $t_n=8.90$ Gyr and Age$(t_n)= 5.11$ Gyr.
Therefore, on average the model with the \citet{chabrier2003} IMF predicts older objects and with a smaller stellar mass content.

\end{document}